\documentclass[acmsmall,screen, languages={Chinese, English}]{acmart}

\usepackage{CJKutf8}

\AtBeginDocument{%
  \providecommand\BibTeX{{%
    \normalfont B\kern-0.5em{\scshape i\kern-0.25em b}\kern-0.8em\TeX}}}

\setcopyright{acmlicensed}
\copyrightyear{2018}
\acmYear{2018}
\acmDOI{XXXXXXX.XXXXXXX}

\acmConference[Conference acronym 'XX]{ACM CSCW}{Jan 03--05,
  2024}{Woodstock, NY}
\acmISBN{978-1-4503-XXXX-X/18/06}

\usepackage{graphicx}
\usepackage{subcaption}
\usepackage{xcolor}
\usepackage{colortbl}

\begin{document}


  \title[ICT-mediated Remote Education in Rural China]{\textit{Who Changed the Destiny of Rural Students, and How?}: Unpacking ICT-Mediated Remote Education in Rural China}


\author{Yuling Sun}
\affiliation{%
  \institution{East China Normal University}
  \country{China}}
\email{ylsun@cs.ecnu.edu.cn}

\author{Xiuqi Zhu}
\affiliation{%
  \institution{Northeastern University}
  \country{USA}}

\author{Xiaomu Zhou}
\affiliation{%
  \institution{Northeastern University}
  \country{USA}}
\email{xi.zhou@northeastern.edu}

\author{Bingsheng Yao}
\affiliation{%
  \institution{Rensselear Polytechnic Institute}
  \country{USA}}
\email{yaob@rpi.edu}

\author{Kai Zhang}
\affiliation{%
  \institution{East China Normal University}
  \country{China}}

\author{Dakuo Wang}
\affiliation{%
  \institution{Northeastern University}
  \country{USA}}
\email{d.wang@northeastern.edu}

\author{Jiaju Chen}
\affiliation{%
  \institution{East China Normal University}
  \country{China}}
\email{10205102450@stu.ecnu.edu.cn}

\author{Liang He}
\authornote{Corresponding author}
\affiliation{%
  \institution{East China Normal University}
  \country{China}}

\renewcommand{\shortauthors}{Yuling Sun, et al.}

\begin{abstract}

The proliferation of Information and Communication Technologies (ICTs) has shown great promise in addressing educational challenges facing rural areas. However, the complex rural context poses significant challenges to the effective utilization of these technologies. This paper examines the empirical integration of live-streaming-based remote classrooms (LSRC) through a qualitative study in rural China. 
Our findings suggest that while LSRC enables rural students equal access to high-quality educational resources, its practical integration faces numerous challenges. In particular, we emphasize the crucial role of local teachers in addressing these challenges, ultimately achieving the desired improvement of students’ learning outcomes. We also examine the impact of LSRC on the original rural education ecosystem. 
Building upon our findings, we call for a reconsideration of interaction paradigms and evaluation systems of ICT-mediated rural education, emphasizing the significance of rural teachers. We conclude by discussing the implications for future ICT-mediated technology interventions in rural settings.

\end{abstract}




\maketitle

\section{Introduction}
The huge educational inequality between regions has been a major concern around the world \cite{attewell2010growing, alvaredo2018world}, especially in developing and underdeveloped countries \cite{bing2008education}.  
As Information and Communication Technologies (ICT) and Artificial Intelligence (AI) become increasingly accessible and available around the world, 
an increasing number of practitioners and organizations have been working on the introduction of educational technology interventions to the low-income and rural areas, intending to address educational inequality issues and to improve the quality of education there \cite{extance2023chatgpt, agbedahin2019sustainable,unesco}.

\textit{ICT-mediated remote education}, which refers to the use of ICTs to deliver high-quality educational content and resources in synchronous \cite{panyajamorn2022using} or asynchronous \cite{guo2022remote} ways, has been increasingly deployed to students in economically underdeveloped, rural, and isolated areas. Recent studies have reported that such an ICT-mediated education paradigm can create a virtual bridge and enable students in disadvantaged settings to benefit from the expertise and resources available in advantaged settings \cite{hohlfeld2017examination, masino2016works, khan2019mitigating}.
Given its tremendous promise and potential to address the educational inequality,
this paradigm has received significant attention globally. Many governments, such as the U.S. \cite{USk12eduplatform}, China \cite{yikuaiping, Chinak12eduplatform}, and India \cite{Indiak12eduplatform}, are deploying it to address their educational inequality issues in rural areas.

Aside from the promising future,
a large body of literature has also noted the cumbersome usability and effectiveness of educational technologies in real-world education scenarios. Studies, for instance, have pointed out that students often encounter various practical challenges in adopting educational technologies, such as limited attention span and sense of engagement \cite{zheng2015understanding, chen2021learning, ramsey2016Preserving}, lack of interaction with teachers and peers \cite{Group2015Salimian, chen2021learning}, etc.
For teachers, the integration of technology interventions often brings certain challenges to their actual teaching process and agency \cite{chen2021Was, varanasi2019teachers}, needing them to reconfigure their original teaching practices to deliver online teaching for better student engagement and learning performance \cite{chen2021Was}. 
Given these challenges, there is a general consensus that social-technical context must be considered when developing and implementing educational technologies \cite{selwyn2010looking, chen2021Was, ng2019shifting}, and evaluation of these technologies in a real-world education context is essential for identifying potential issues that inhibit successful implementation \cite{taha2023impact, sulecio2018avoiding}.

In rural and isolated regions, these practical situations become even more challenging and complicated, due to the often limited digital infrastructure \cite{salemink2017rural}, technical literacy of students and teachers \cite{chaklader2013bridging, hardy2019rural}, diverse student backgrounds, learning styles, and needs \cite{de2014reasons, johnson2007rural}, as well as specific socio-cultural and socio-political background and educational awareness \cite{johnson2007rural, guzman2021dropout}. 
Further, the widespread lack of digital literacy in rural areas \cite{chaklader2013bridging, hardy2019rural} significantly impacts the acceptance of local students and teachers in using educational technologies (also known as the 'digital divide' \cite{chaklader2013bridging}), which potentially lead to further educational inequality \cite{sulecio2018avoiding}. Given the known challenges facing ICT-mediated education paradigm, as well as the more complicated situation of rural areas, it is critical to examine how ICT-mediated education paradigm is integrated and used (or not) in real-world rural education settings, and how stakeholders (students and teachers) experience and perceive this paradigm. With technology interventions becoming increasingly popular in rural educational settings, understanding this question can shed light on how we can design to better fit technology interventions into rural education scenarios, with increased system uptake and education outcomes. 

To this end, our study conducted a qualitative study of ICT-mediated remote education in rural China. We chose China to be the research site because of its huge rural population \cite{liu2017rural}, the challenging rural education situation \cite{yue2018rural}, and the national strategy of ICTs mediated Collaborative Education Quality Enhancement Initiative\footnote{Collaborative Education Quality Enhancement Initiative 
("\textit{xie tong ti zhi}", \begin{CJK}{UTF8}{gbsn} 
协同提质
\end{CJK}) 
refers to leveraging ICTs to create a connection between the advantaged and disadvantaged areas and enhance the quality of rural education through collaborative efforts.} \cite{xietongtizhi}. 
In recent years, a series of ICT for rural education projects have been promoted, which offer us the opportunity to closely examine these ICT interventions in an empirical rural education setting. 
We specifically examine “\textit{One Screen}” \cite{yikuaiping}, the most representative ICT-mediated remote education initiative tailored for rural students in China. 
The initiative introduces live-streaming-based remote classrooms (i.e., “one screen”) as well as other accompanying devices to rural K-12 schools. 
These devices are connected to a well-resourced, prestigious, and top-ranked school in metropolitan cities (e.g., Chengdu, Sichuan province). 
Currently, \textit{One Screen} initiative has already deployed to over a million students in rural China \cite{yikuaipingdata} (see more background information in Section~\ref{sec:methods}). Through a three-week field observation of such live-streaming-based classes in a rural school and an interview study with 15 stakeholders (14 local teachers and 1 project manager), this paper aims to the following research questions: 

\begin{enumerate}
    \item \textit{How does live-streaming-based remote class education paradigm work in practice in a real-world rural educational setting?}  
    \item \textit{How do the local students and teachers use, experience and perceive this new education paradigm?}
    \item \textit{How does this new education paradigm impact the original educational ecosystem in rural areas?}
\end{enumerate}

Our findings suggest that while live-streaming classes allow rural students equally access high-quality educational resources, its practical implementation faces numerous challenges: students lacking the necessary knowledge foundation and skills to adapt and absorb high-quality educational content and the limitation of live-streaming classes in maintaining students' engagement. More importantly, our findings emphasize the crucial role local teachers play in the original education system in addressing these challenges. 
They reconfigure their identity from content creators to content deliverers and enable students to better absorb and comprehend this content through a series of efforts, ultimately improving students' learning outcomes. 
We also analyze the impact of this technological intervention on local stakeholders, especially teachers, in terms of workload, self-identity, and personal development. 

Our paper makes the following contributions to CSCW: (1) We offer a qualitative, empirical understanding of how \textit{One Screen}, a kind of live-streaming-based remote classroom, is integrated and works in the real-world rural school, elaborating the practical procedure, practices, encountered challenges and underlying reasons. 
Given the development history of over a decade \cite{yikuaipingdata} and a user base exceeding millions, we believe that the experiences and lessons learned from \textit{One Screen} initiative can provide invaluable empirical data for the community.
(2) We expand existing knowledge of technology interventions in rural education settings by identifying the significant, irreplaceable role of local teachers in filling the practical challenges facing technology interventions, calling for attention to the role of local teachers in enabling students to better absorb and comprehend high-quality education content, maintaining students' attention and engagement in class, and promoting students' active thinking.  
(3) We relate our findings to current knowledge of the impact of technology interventions on education equality and equity in rural areas, highlighting more flexible and adaptive technological interventions tailored to the specific socio-technological context in rural areas. Finally, (4) we offer design opportunities tailored to the live-streaming-based education paradigm with the aim of enhancing the practical efficacy and effectiveness of this paradigm and alleviating the workload of local teachers.

\section{Related Work}

Our study is situated in the research of ICT-mediated remote education, particularly in rural and isolated areas, and its practical effectiveness and impacts.
In this section, we first summarize prior research that 1) explores existing technology-mediated remote education interventions, and 2) particularly in rural and isolated areas. Following that, we turn to the impact of technological interventions and present 3) the existing discussion of the impact of technological interventions on teachers’ roles and work, and 4) educational equality and equity. These four lines of research set up the background for understanding the contributions of our study.

\subsection{ICT-Mediated Remote Education}

In terms of temporal synchronization, technology-mediated remote education can be generally divided into asynchronous and synchronous learning. Asynchronous learning mainly includes Massive Open Online Courses (MOOCs) \cite{zheng2015understanding, hew2014students} and Online Degree Courses \cite{boton2015minimizing}. In such a paradigm, course videos are prerecorded and uploaded to online platforms by teachers, and learners can flexibly access these courses according to their own situations. Synchronous learning provides a more straightforward learning environment wherein learners can interact with teachers in real time. A typical synchronous learning format is live-streaming-based learning (LS learning), with two major categories: video conference \cite{chen2021learning} and live streaming \cite{chen2021Was}.
 
Prior works in computer-supported cooperative work (CSCW) and related fields have examined stakeholders’ motivation, experiences and encountered challenges when engaging into various forms of remote education (e.g., \cite{chen2021learning, zheng2015understanding, chen2021Was, song2004improving, gillett2017challenges}). \citet{zheng2015understanding}, for instance, concluded four types of students' motivations for learning by MOOCs, including fulfilling learning needs, preparing for the future, satisfying curiosity, and connecting with others. To characterize the online community of learners, \citet{sun2018community} conducted a large-scale survey among an online learning platform, and revealed five factors affecting learners' online experience, including shared identity, interpersonal friendship, identity regulation, coordination and social support. Research on LS learning has explored users’ experiences and challenges when involving LS learning \cite{chen2021learning, chen2021Was} as well as various specific knowledge learning practices, such as fitness exercise instruction \cite{guo2022s}, program training \cite{faas2018watch} and so on. 
 
Compared to these remote education paradigms, the ICT-mediated remote classroom our study focused on (also referred to as blended synchronous learning \cite{bower2015design} or hybrid flexible teaching \cite{detyna2023hybrid}), is a kind of digital-physical hybrid education paradigm, which combines elements of both digital and physical learning modalities to create a comprehensive and blended educational experience. To be specific, it often utilizes live-streaming to deliver real-time teaching videos and related educational materials from the classroom in an urban or well-resourced setting to the classroom in a rural or remote area. Students in rural and remote areas have access to and learn high-quality educational content in their physical classrooms.

In CSCW, although there has been surging interest in research on both asynchronous (e.g., \cite{chen2021Was, chen2021learning}) and synchronous (e.g., \cite{zheng2015understanding, sun2018community} remote education, with exception of \citet{guo2022remote},
limited attention has been paid to this increasing digital-physical hybrid education paradigm. 
This research gap is critical because researchers need an in-depth understanding of how to appropriately implement this technology intervention to increase the practical educational outcome as well as the education experience of intended users (e.g., teachers and students).
We aim to fill this gap by enriching empirical understandings of this ICT-mediated hybrid education paradigm, in particular in rural and isolated settings. 

\subsection{ICT-Mediated Rural Education}
Rural education is known to play a key role in socio-economic development \cite{schultz1960capital, brown2002education}. How to improve the educational quality in rural and isolated areas has consistently been the focal point of attention among academic, governmental bodies, and nongovernmental organizations \cite{UN_EDU, extance2023chatgpt}. In recent years, a growing number of practitioners and organizations believe the increasingly accessible and affordable ICT and AI technologies can potentially address the challenges rural and isolated education faces, as they have become increasingly powerful and are capable of building powerful and sophisticated learning environments \cite{smeets2005does, extance2023chatgpt}, supporting multiple learning and teaching needs \cite{chen2021Was}, and transforming pedagogical practices in the classroom \cite{guo2022remote}. Many governments and organizations have launched ICT-assisted education initiatives targeting rural and isolated areas, such as hole in the wall \cite{mitra2005acquisition}, One Laptop per Child (OLPC) project \cite{ames2016learning}, South Africa's Technology for Rural Education project (TECH4RED) \cite{techred}, and China’s One Screen project \cite{yikuaiping}. 

In CSCW, a large body of research has been working to design and deploy technology interventions that aim to improve the quality of education in low-income rural and isolated schools \cite{arawjo2019computing, makalima2023co, khan2019mitigating}. For instance, \citet{makalima2023co} co-design an integrated digital education portal with educational stakeholders to improve their access to educational resources, and \citet{khan2019mitigating} design mobile applications for bridging the educational gaps between urban and rural learning environments. Some studies have conducted targeted technology design specifically addressing the distinctive features of rural education \cite{patel2015design, cidota2020remote}. For instance, \citet{patel2015design} explored design opportunities for learning aids in multi-grade schools in rural India, and \citet{cidota2020remote} used Augmented Reality (AR) based technical system to promote virtual co-location, with the aim of reducing the risk of early school dropout in rural areas.
Furthermore, some technologies and tools were designed to support rural teachers’ professional development and capacity training \cite{botha2015icts, menon2017technology}, and assist their daily teaching practices \cite{guo2022remote, liu2018virtual, varanasi2019teachers}.
In addition, researchers have also designed technologies tailored to specific functions, aiming to support the learning and training of rural children’s specific skills which they often lacked the effective means to learn, such as spoken English \cite{larson2013want}, programming \cite{mo2015computer}, technology and AI literacy \cite{arawjo2019computing, vandenberg2023leveraging, jonas2022digital}.

 
These research efforts have greatly advanced the development of ICT-mediated rural education. 
Yet, increasing researchers have pointed out that importing these technological interventions and making them fit into rural educational settings were often much more challenging, even resulting in failure \cite{ford2010pragmatic, bladergroen2012educator, were2011bridging}, due to limited digital infrastructure \cite{were2011bridging}, technical literacy of students and teachers in rural and isolated areas \cite{ford2010pragmatic, gudmundsdottir2010does}, diverse student backgrounds, learning styles, and needs \cite{salemink2017rural, de2014reasons}, as well as specific socio-cultural and socio-political background \cite{wani2022hartal} and educational awareness \cite{de2014reasons}. Researchers, therefore, have articulated a need to carefully examine the implementation and development of technological interventions in rural and isolated educational settings. Our study adds to these ongoing efforts to deeply understand live-streaming-based remote classroom education in rural China.
 
\subsection{ICT-Mediated Education and Rural Teachers}


As technological interventions are increasingly introduced into traditional classroom education, 
CSCW research has also paid attention to designing and examining the technology interventions from teachers' perspectives \cite{varanasi2019teachers, leach2006deep,guo2022remote, zhang2023case}.
For instance, through a specific case study of a teacher-focused intervention in India, \citet{varanasi2019teachers} show how the introduction of technology interventions causes teachers to reconfigure their work practices, including lesson preparation, in-classroom teaching practices, bureaucratic work processes, and post-teaching feedback mechanisms. \citet{guo2022remote} examine China’s co-teaching classroom program, and find that remote teachers’ high-quality classes directly impacted local teachers and students, while local teachers were also actively involved in making indirect impacts on their students by deeply coordinating with remote teachers and adapting the resources offered by the remote teachers.

In particular, existing studies have provided strong evidence that teachers play significant roles when introducing technological interventions into rural educational settings. Researchers claimed that teachers need to develop technological pedagogical knowledge (TPK) and technological content knowledge (TCK), which are critical for them to successfully integrate technology use into teaching \citet{varanasi2019teachers, koehler2009technological}. 
\citet{koehler2009technological} proposed a Technological Pedagogical Content Knowledge (TPACK) framework, considering the amorphous nature of technology and encouraging teachers to adopt technology to improve their content and pedagogical capacities while avoiding technocentric approaches \cite{schmidt2009technological}. 

Yet, researchers have pointed out that while ICT could potentially help address challenges such as inadequate resources and under-qualified educators in practical rural settings, it is in resource-constrained environments that teachers have limited TPK and TCK. They encountered a series of challenges in integrating the use of ICTs into teaching and learning. For instance, it has been largely reported that most teachers had inadequate ICT and pedagogical competencies for effective integration of ICT into their work \cite{botha2015icts, rambrij2018technology, grad2020examination}. To address this challenge, a number of recent technology-based teacher professionalization efforts have focused on supporting and building teacher capacities in the field and meeting the challenges posed by technology-intervented rural teaching and learning environments \cite{koehler2009technological, botha2015teacher}. 

Moving forward, recent studies have also indicated that many existing technology interventions, such as hole in the wall \cite{mitra2005acquisition}, OLPC \cite{ames2016learning}, and the FATIH program \cite{isci2015use}, have taken technology-centric approaches, without providing teachers with a central role (completely or partially) in the deployment of these technology interventions. That is, teachers are viewed as passive actors, and their teaching practices are largely disregarded when designing interventions. This gap is critical because teachers often play a significant role in delivering technology interventions to students \cite{varanasi2019teachers, koehler2009technological}, and their experiences also directly influence the practical effectiveness and efficacy of these technological interventions \cite{guo2022remote}. 
Consequently, researchers have suggested a shift from technology-centric intervention approaches towards creating technology-enhanced education environments that place teachers at the center of the intervention, and provide strong support and capacity building for teachers, which would be helpful for leading to more meaningful integration of technology in classrooms \cite{varanasi2019teachers, koehler2009technological, gerard2011professional}. Our research builds on this literature by contributing a qualitative empirical study that examines how teachers in Chinese rural schools reconfigure their everyday work practices around live-streaming-based remote classrooms and how this kind of technological intervention impacts their teaching work.


 
\subsection{Educational Equality and Equity, and Technology}
Education equality and equity are two essential and crucial terms in education \cite{rachid2022quality, ainscow2020promoting}. They look similar but actually have vastly different meanings. Equality in education often means that every student receives the same resources and opportunities \cite{twoTerms}. It is beneficial but often does not address specific needs. Instead, education equity refers to that each student receives what they need to develop to their full academic and social potential \cite{twoTerms2}. 
Both equity and equality are important to ensure the best possible outcomes for students \cite{twoTerms2}. 
Recent international policy documents argue that they should be seen as principles that inform all aspects of educational policy \cite{unesco2017guide}.

Equality and equity have been a matter of significant concern in rural education settings. 
In terms of educational equality, research has demonstrated that students in rural areas often have limited quality teachers \cite{goldhaber2015uneven, podolsky2016solving, burroughs2019teacher}, limited access to quality education resource \cite{makalima2023co, unesco2016guide}, and limited infrastructures \cite{chaklader2013bridging, were2011bridging}, which caused crucial educational inequality. In terms of education equity, significant disparities in fundamental demographic characteristics such as gender \cite{pappu2021gender, anderson2015digital}, ethnicity \cite{webb2016unfulfilled}, and diverse student features and needs \cite{salemink2017rural, guzman2021dropout} bring significant challenges of educational equity in rural areas. The limited availability of resources \cite{ruralequity} and an imperfect social security system \cite{shehuibaozhangrural} further exacerbate these challenges.

Recently, the development of ICT and AIs has enabled education reform in rural areas, with tremendous promise for addressing issues of educational equality and equity in rural areas. They offer an affordable and accessible means, enabling more rural students to equally access educational resources \cite{mitra2005acquisition, ames2016learning, techred, yikuaiping}; Meanwhile, various technology designs tailored for rural teachers (e.g., \cite{botha2015icts, menon2017technology, guo2022remote, liu2018virtual, varanasi2019teachers}) and students \cite{robinson2008using, makalima2023co, khan2019mitigating, patel2015design, cidota2020remote} are considered crucial tools and catalyst in addressing equity challenges facing rural education.
Yet, an increasing number of researchers have noted that imbalanced or improperly utilized ICT resources \cite{yang2018promoting}, as well as the differences in digital literacy among rural students \cite{chaklader2013bridging, anderson2015digital} may exacerbate, rather than contract existing inequalities \cite{sulecio2018avoiding}. Researchers articulate an urgent need to prudently evaluate the impact of technology interventions on educational equity \cite{sulecio2018avoiding, yang2018promoting}. Our work adds to this literature by deeply examining the practical impact of ICT-mediated remote education on rural education quality, equality, and equity.

\section{Methods}
\label{sec:methods}
The goal of our study is to gain an in-depth empirical understanding of ICT-mediated remote classes, in particular, live-streaming-based remote classes (LSRC) in real-world rural settings.
We are particularly interested in 1) how LSRC is integrated and works in the real-world rural class, 2) how local stakeholders (teachers and students) use, experience, and perceive this paradigm, and 3) how it impacts their practical teaching and learning behaviors. 
To investigate these questions, we grounded our study in the context of one specific rural school in remote China, which had adopted LSRC initiative (i.e., “one screen”) for more than four years. We conducted a qualitative study, including a three-week observation of teaching practices in practical LSRC, interviews with 14 local teachers and 1 project manager, and informal conversations with a number of students. The first author, a native Mandarin speaker, conducted the data collection, and three coauthors analyzed the collected data together. 
We now elaborate on the research background as well as details on the data collection and analysis processes. 

\subsection{Research Background: “One Screen” Initiative in China}
As the most populous developing country globally, rural education has always been a crucial societal issue in China \cite{yue2018rural}. To address educational inequalities and promote educational development in rural and ethnic minority areas, the central government of China, along with the education authorities at various levels, educational research institutions and organizations, as well as social welfare organizations, have actively sought solutions. 
With the rapid development of ICT technologies, educational institutions in China have begun exploring the ICT-mediated remote classroom paradigm as a solution \cite{yikuaiping}, in which ICT technologies are utilized to bridge the geographical gap and connect two distinct learning environments - often one in an urban or well-resourced setting with potentially better educational infrastructure and resources, and the other in a rural or remote area where such resources may be limited. ICT here serves as the mediator, facilitating the transfer of educational materials, including instructional images, sounds, text, multimedia animations, and video materials, from the urban classroom to the rural one in real-time.

With the tremendous promise of addressing educational inequalities by providing students in remote or under-served areas with access to high-quality education, the ICT-mediated remote classroom paradigm has rapidly gained widespread attention. In 2005, the Ministry of Education of China initiated the "National Basic Education Alliance\footnote{https://m.g12e.org/}", a non-profit organization dedicated to promoting this model nationwide. 
Driven by the Alliance and strong government supports \cite{support1, support2}, this ICT-mediated remote classroom paradigm and similar remote education initiatives have flourished in China. 
In 2018, the news titled "This Screen Might Change Their Destiny \cite{destiny}" drew attention to the impact of this model on rural children, spreading widely in Chinese public discourse and sparking extensive discussions on educational opportunities and fairness in remote areas. Motivated by this, NetEase\footnote{NetEase, Inc. is a leading Chinese technology company that provides a diverse range of internet services, including online gaming, education, e-commerce, advertising, music streaming, and more.} launched the "One Screen" philanthropic project \cite{yikuaiping}\footnote{https://gongyi.163.com/yikuaiping/}, with primary aim of introducing more live-streaming-based remote classroom (i.e., “screen”) as well as other intelligent software and hardware platforms to remote schools, and providing equal educational opportunities to students there. According to the report from the Counsellors' Office of the State Council of China \cite{yikuaipingdata}, the "One Screen" and similar educational projects have covered nearly 10,000 rural or remote schools nationwide, benefiting over a million students. The long-term and large-scale deployment of "One Screen" initiative in China provides us with unique opportunities to deeply examine the effectiveness and impact of ICT-mediated rural classrooms in real-world settings.

\subsection{Data Collection: Participatory Observation}
We began with a three-week observation in XD School (anonymous name), a high school in the rural area of Yunnan province, China, that had adopted the “One Screen” paradigm for more than four years.
The purpose of the observation was to gain a thorough understanding of how LSRC was integrated and worked in the real-world rural classroom (RQ1) while shaping the subsequent interviews.

\subsubsection{Recruitment and Criteria}
XD school was recommended by CD High School (anonymous name), the responsible institution of one well-known K-12 LSRC initiative in China.
The institution where the first author was affiliated was an educational research organization, that had collaboration with CD High School. Building on the existing collaborative relationship, we reached out to the leader of the LSRC project at CD High School, disclosing our intention to do research about LSRC in rural settings, requirements (including long-term observation and interviews with stakeholders), research ethical approval, and commitments to data privacy and safety. 
The project leader was supportive of our research and appreciated our interests and attention to them. They also hoped our study could help them better design future LSRC and improve stakeholders’ using experiences and learning outcomes.
 
With the criteria of recruiting schools 1) who had adopted LSRC for at least two years and 2) located in rural areas, the project leader of CD High School recommended three candidate schools for us.
We then reached out to these three schools, seeking permission to conduct research. Two schools agreed to be studied, and one declined with the reason of “the current project is still in its early stages and not suitable for research”.
We finally chose XD school as our research site, because it was located in a remote and impoverished rural area of Yunnan province, China, and a significant portion of local residents there belong to ethnic minorities. We considered these features to make it an ideal research site for our research.

\subsubsection{Observation}
Observation primarily took place in classrooms and teachers' offices. After receiving permission from XD school, the first author started the three-week observation (seven days per week 7 am to 11 pm per day, following school’s daily schedule). 
The observation was primarily conducted in the first and second grades of senior high school. We didn’t observe the third-grade classroom because the school principal suggested that both teachers and students in the third grade of senior high school faced significant teaching and learning pressure, and it was not recommended for us to conduct research there. 

According to the introduction by the principal, the LSRC included 9 subjects for the college entrance examination (i.e., Chinese, Mathematics, English, Politics, Geography, History, Physics, Chemistry, Biology), and other courses (such as Information Technology, Psychology, Physical Education, etc.) did not involve into LSRC initiative. 
At the beginning, based on the schedule provided by the principal, we randomly selected classes (including both LSRC and non-LSRC) for observation to gain a general understanding of the entire teaching process and environment in the school. The first author positioned herself at the back of the classrooms, observing and recording the entire class process, encompassing pre-class, in-class, and post-class activities, without disrupting the teaching process.

After gaining an overall understanding, she then selectively chose LSRC to deeply understand the detailed utilization process of LSRC, as well as teachers' and students’ teaching and learning processes and practices in such kinds of classes. 
During breaks, she randomly communicated and interacted with students to understand their class experiences, feelings, encountered issues, and solutions.
Handwritten field notes, photographs, and videos were used to record data. 
In addition, employing the behavioral observation method \cite{bakeman2012behavioral}, we observed and recorded the daily work processes and activities of teachers.

After three weeks of observation, we observed a total of 64 LSRCs (covering all live-streaming subjects for the first and second grades) and 24 non-LSRC classes (including self-study sessions, lab classes, psychology classes, etc.), and six teachers’ daily work practices (three ones of science courses and three ones of humanities courses). Through these, we gained comprehensive insights into how LSRC was integrated and worked in the practical rural class, which also shaped our subsequent interview questions.

\subsection{Data Collection: Semi-structured Interview}
To deeply examine how local stakeholders experienced and perceived LSRC paradigm (RQ2), and how this kind of technological intervention impacted their practical teaching and learning behaviors (RQ3), and help address certain questions that we could not during observation, we then conducted in-depth interviews with 14 teachers who had integrated the LSRC into their own class, and 1 vice-principal of XD school who was in charge of LSRC project). 
We did not conduct in-depth interviews with local students because, during the observation process, we found that local students face tremendous academic pressure, studying seven days a week from 7 a.m. to 11 p.m. daily. We didn't want to interfere with their study time. Therefore, data about students' experiences, perceptions, and impacts were primarily gathered through 1) brief conversations with students during break times and 2) insights from teachers.

\subsubsection{Recruitment and Participants}
Our participants were primarily recruited during the observation, with the criteria of 1) teachers who had integrated LSRC into their own classes and 2) who had covered a diversity of subjects and different grades. 
During the observation, the first author actively reached out to teachers of LSRC, disclosed our intention of conducting research and recruited them for our interviews. We eventually recruited 14 teachers. In addition, we also recruited the vice-principal of XD school who was in charge of the LSRC project to our interview study. The basic information of our participants are shown in Table 1. We used P1-P14 to represent 14 interviewed teachers, and M15 to represent the vice-principal.

\begin{table}[ht]
\centering
\caption{Summary of participants. }
\label{tab:participants}
\begin{tabular}{llllp{0.3in}llll}
\hline
\rowcolor[HTML]{C0C0C0} 
\textbf{ID} & \textbf{Gender} & \textbf{Grade} & \textbf{Subjects} & \textbf{} & \textbf{ID} & \textbf{Gender} & \textbf{Grade} & \textbf{Subjects} \\ \hline
P1          & Male            & 2nd            & Physics           &           & P9          & Female          & 2nd            & Chemistry         \\
P2          & Female          & 1st            & Mathematics       &           & P10         & Female          & 2nd            & Geography         \\
P3          & Male            & 2nd            & Chinese           &           & P11         & Male            & 1st            & Biology           \\
P4          & Male            & 1st            & Chinese           &           & P12         & Female          & 2nd            & Politics          \\
P5          & Female          & 2nd            & Mathematics       &           & P13         & Female          & 1st            & History           \\
P6          & Male            & 2nd            & Biology           &           & P14         & Male            & 1st            & Physics           \\
P7          & Female          & 1st            & English           &           & M15         & Male            &                & Vice-Principal    \\
P8          & Female          & 2nd            & English           &           &             &                 &                &                   \\ \hline
\end{tabular}
\end{table}

Of the 14 teachers, 6 were male and 8 were female; 8 were from the first year of high school and 6 were from the second year. Their classes covered 9 subjects, including Chinese, Mathematics, English, Politics, Geography, History, Physics, Chemistry and Biology.

\subsubsection{Semi-structured Interview}
In our interviews with the teachers, we asked questions about 1) their personal history of adopting and using processes when integrating LSRC into their own classes, 2) benefits and challenges they experienced, 3) reactions towards the challenging situation, if any, 4) the changes LSRC brought to them, 5) perceived functions and impacts of this paradigm to rural students, themselves and rural education, and 6) their attitudes and perceptions to future technology mediated educational interventions. Interviews with the vice-principal included questions similar to those with teachers, but from the educational management’s perspective. 
All questions were designed to be general so that they could be inclusive and encourage interviewees to share various levels of experiences and perceptions. 
We would probe for more details and concrete examples when participants shared more interesting points or prior experiences.
In addition, we also had a range of tailored questions targeting each specific teacher based on our earlier observations, such as motivation, purpose, and reason behind a specific teaching behavior we observed. 

All interviews were conducted by the first author in person at school offices; each lasted for 30-40 minutes. With participants’ permission, all interviews were conducted in Mandarin, audio recorded, and were later transcribed in Chinese verbatim for analysis.

\subsection{Data Analysis}
We followed an inductive process for analysis \cite{glaser1978theoretical}. During the open coding phase, three researchers started independently coding and analyzing while the data was being collected and developed an initial set of codes to exhaust the data.
This process was iterated by generating codes from the collected data, classifying different codes into similar themes, regularly discussing with each other to ensure reliability, and checking and elaborating these codes and themes as new data were obtained.
The data collection process stopped once all the core variables were saturated. 

By the end of the open coding phase, we generated an initial set of codes, capturing teachers’ teaching procedures and actions (including their course preparation, guiding students in pre-class preparation, Assisted lecturing, assignment grading, post-class tutoring, etc.), their communications and collaborations with remote teachers, personal experiences, benefits and challenges of this paradigm to local teachers and students in their minds, their reactions to the encountered challenges, their concerns, perception and expectations related to LSRC and the other technological interventions in rural educational setting.
Based on these preliminary themes, we then iterated between the contextualized meanings of emergent themes and reviewed related literature to uncover their connections.
Through this process, all data and concepts gradually converged on four interrelated themes: 
(1) the practical procedure and collaborations of LSRC in rural settings, (2) the challenges when integrating LSRC into the rural class, (3) teachers’ efforts in addressing these challenges, and (4) the impact of LSRC on rural teachers and students. 
We present the full details of our findings in the next section, using representative quotes, which were then translated into English. 
Considering that we took rural settings as the primary research sites, focusing on the practical situation of LSRC in rural settings, we used ‘remote X’ to present the stakeholders and environments in the urban settings (i.e., CD school), and ‘local X’ to present ones in rural settings (i.e., XD school) in the following Finding sections.

\subsection{Ethical Considerations}
We obtained research ethical approval from the Ethics Committee of the authors' institution to conduct all the procedures of our study involving human subjects. 
All researchers of this study received formal ethics training and were granted official certificates from ethics committees in their institutes. 
Throughout the research process, we took careful steps to protect research subjects' rights and privacy. 
Before starting, we informed all participants our intention and background information, our promise that all collected data would be used only in this study, and got their permission. 
All collected data were used in an anonymous approach(i.e., there was no link between the collected data and an individual subject).

\section{Findings}
\label{sec:findings}
Our findings are organized around the functioning, challenges, resolution processes, and impacts of live streaming-based remote classes (LSRC) in real-world rural settings. We will first present the anticipated procedure of LSRC. As a baseline, we elaborate on the collaboration among stakeholders in dual settings within real-world scenarios (Section~\ref{sec:discussion-roleoflocalteacher}). Following this, we delve into the challenges encountered by this model in a local setting  (Section~\ref{sec:discussion-socialecology}), highlighting how local teachers, through a series of efforts, address these challenges, making this approach effective and beneficial in practical situations (Section~\ref{sec:discussion-qualityequalityequity}). Finally, we report the impacts of LSRC on both local teachers and students (Section~\ref{sec:discussion-designimplications}).

\subsection{Live-Streaming-Based Remote Class (LSRC): Procedure and Collaborations}
\label{sec:findings-procedure}
\begin{figure}
  \centering
 \begin{subfigure}[b]{0.494\textwidth}
    \includegraphics[width=\textwidth]{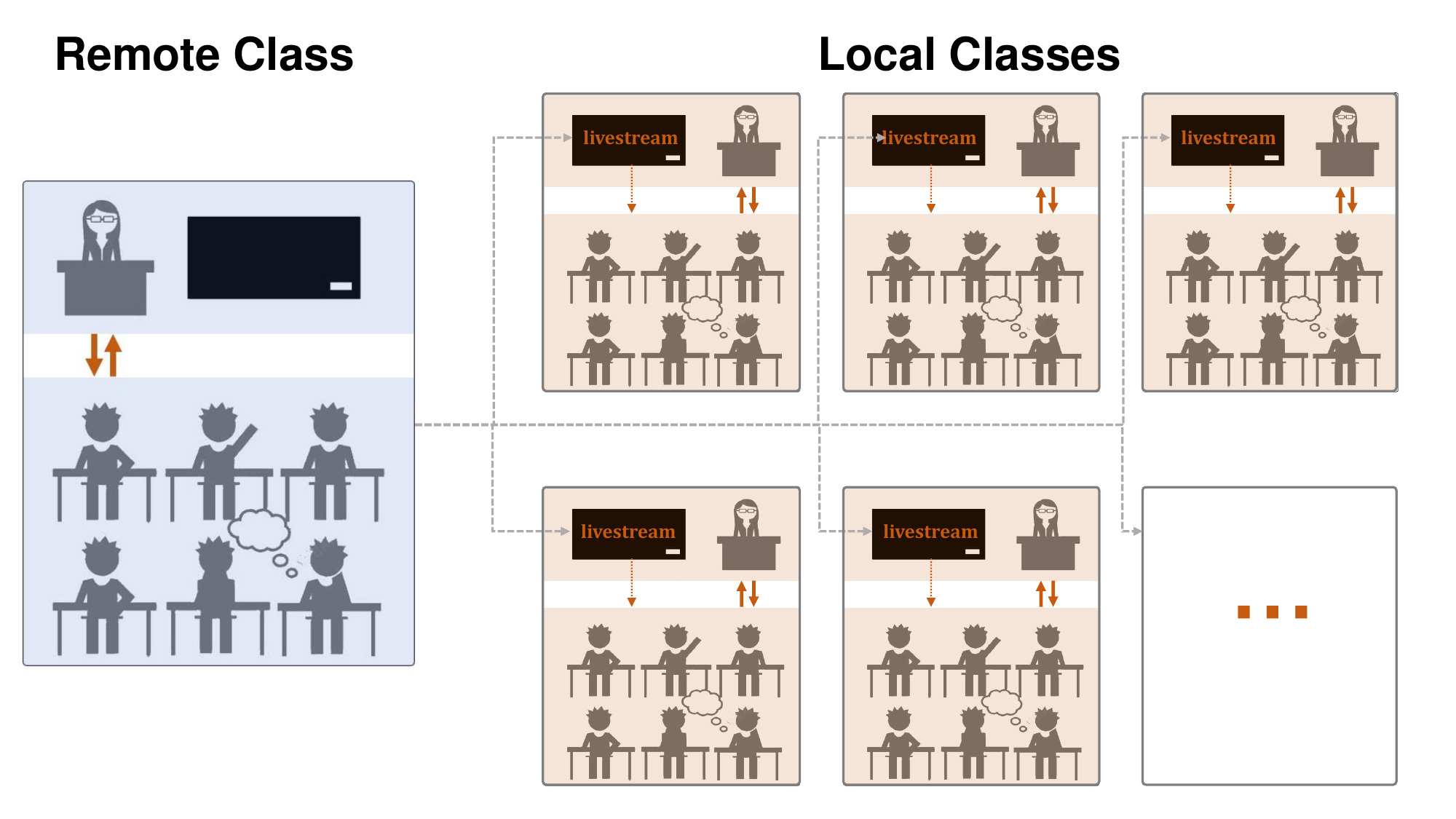}
    \caption{}
    \label{fig:1forAll}
  \end{subfigure}
  \hfill
  \begin{subfigure}[b]{0.494\textwidth}
    \includegraphics[width=\textwidth]{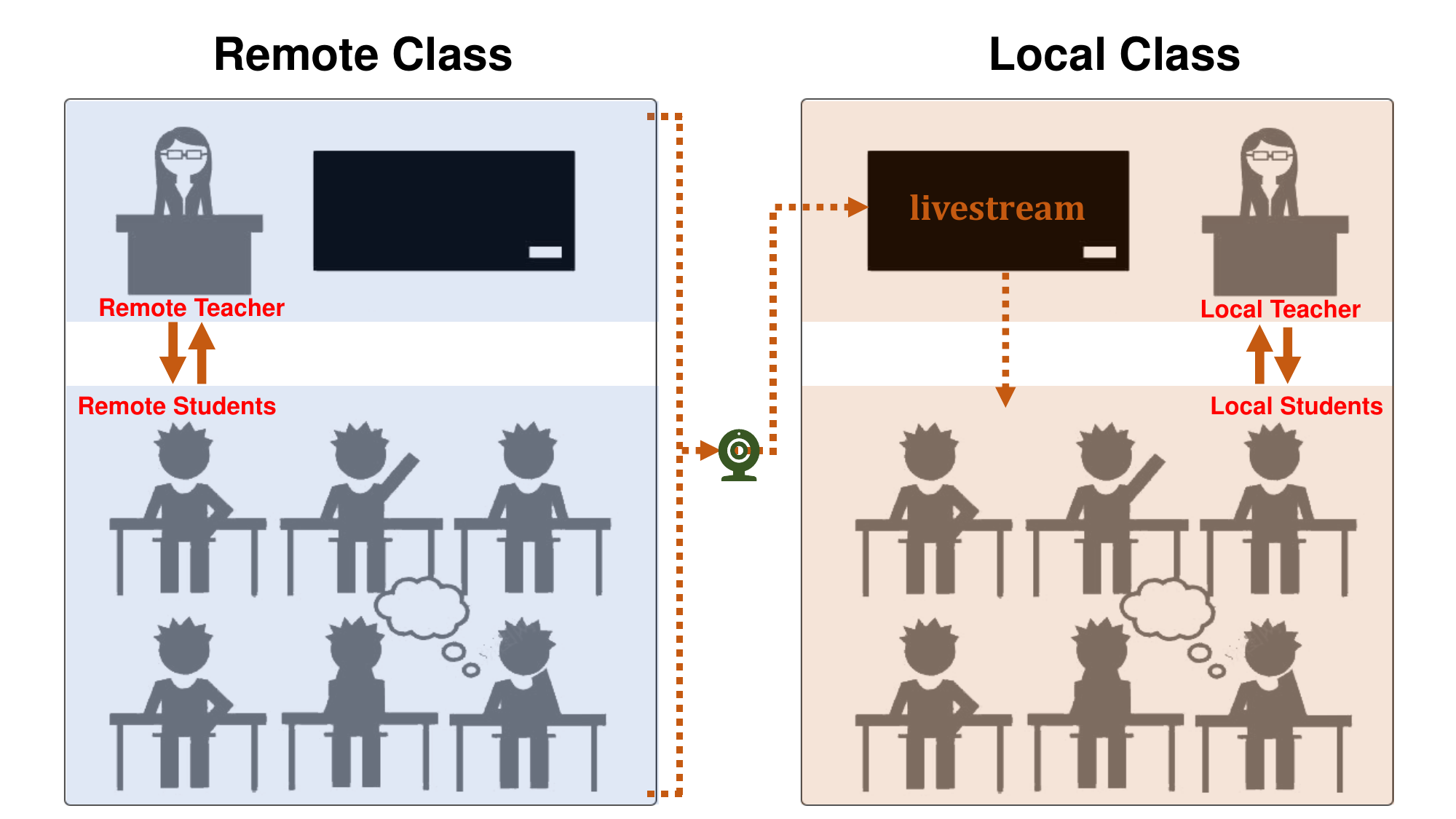}
    \caption{}
    \label{fig:1on1}
  \end{subfigure}

  \caption{The sketches of live-streaming-based remote class (LSRC). (a) is the overview of the entire LSRC initiative, and (b) is a concrete example.}
  \label{fig:overview}
\end{figure}

Figure \ref{fig:overview} illustrated the sketches of LSRC (i.e. One Screen) initiative in rural China. Specifically, when the remote (urban) teacher at CD School conducted classes for their students, all teaching activities were transmitted in real-time through live broadcasts to classrooms in over 300 rural schools (as per M15's description), wherein a local (rural) teacher would provide on-site assistance.

\subsubsection{Procedure of LSRC}
\label{sec:findings-procedure-procedure}
According to M15's introduction, promised LSRC followed the principle of "Four \textbf{S}" between remote and local settings, that is, preparing the lessons \textbf{S}ynchronously, lecturing \textbf{S}ynchronously, with \textbf{S}ame education material and \textbf{S}ame homework.
The general procedure of LSRC (based on our observations) involved: Before each class, remote teachers shared the next day's class materials (e.g., presentations and assignments) with the local teachers; During the class, the remote teacher delivered the lesson, while the local teacher assisted with on-site teaching; After the class, local teachers took responsibility for guiding students through their assignments. Meanwhile, remote and local teams conducted weekly online meetings to collaboratively prepare for the upcoming week's lessons.


\subsubsection{Collaborations and Interactions Between Local and Remote Stakeholders}
\label{sec:findings-procedure-interaction}
Given the promised vision, our study found that the interaction and collaborations between local and remote stakeholders in real-world LSRC scenarios differ somewhat.
Specifically, compared to the findings in \citet{guo2022remote}'s study on remote co-teaching, where a series of interactions and collaborations between remote and local stakeholders have been identified, our study found that in LSRC scenarios we observed, the collaborations between remote and local stakeholders were very limited (as shown in Figure 1-b). Even when some collaborations existed, they tended to be more formalistic and failed to achieve true collaboration in practice. 

\textbf{\textit{Collaborations and Interactions Between Local and Remote Teachers.}}
Based on our observation, the weekly collaboration between local and remote teachers in course preparation was a somewhat formalistic form of collaboration. 
Specifically, in the weekly course co-preparation session we observed, the typical process was that one teacher in CD school and more than one hundred teachers in rural schools joined in an online meeting room and "prepared" the lectures together. Yet, in practice, this scenario was more like the remote (urban) teacher telling or sharing with rural teachers what her lecture content was and what the teaching process and objectives she had designed, rather than all teachers working together to prepare the lecture content. Our participants considered that it was practically almost impossible for rural teachers to participate in the design of the course design because ``\textit{there are too many teachers. Each has his/her own student situations and requirements} (P7).'' Consequently, remote teachers could only prepare the course content according to their own students' situations. 
The only collaboration we observed in this kind of weekly meeting was that the local (rural) teacher informed the remote (urban) teacher through the conferencing software's chat box when there were typos or other minor errors in course content.

\textbf{Collaborations and Interactions Between Local Students and Remote Teachers.} As illustrated in Figure 1-b, the interactions between the remote teachers and local students were limited to one-way communication. That is, the remote teacher gave instructions (e.g., ``\textit{Please take out the draft paper, remote (rural) students as well}'') or asked questions, and the local students followed or answered them. However, the remote teacher wouldn't receive the answer or provide any follow-up feedback. Local students we informally communicated with generally considered the interactions between remote teachers and them to be ``\textit{impossible}'' because ``\textit{many students attend the class online simultaneously}''. 


Instead, our study suggested that most interactions in LSRC paradigm actually happened between local teachers and local students in a two-way means. For instance, when students in local classrooms answered the questions from remote teachers, direct answers and follow-up interactions were provided by local teachers. Further, local teachers also conducted a series of other interactions to support local students' adoption of and understanding of the course content. We now turn our focus to LSRC in rural settings, elaborating on the challenges LSRC paradigm encountered in rural settings (Section~\ref{sec:findings-challenges}), and how teachers reconfigured their teaching practice to address these challenges (Section~\ref{sec:findings-localteacher}), and the impact of LSRC on local teachers and students (Section~\ref{sec:findings-impacts}).

\subsection{Challenges of Integrating Live-Streaming-Based Remote Classroom in Rural Settings}
\label{sec:findings-challenges}
Our study suggested that, integrating LSRC into rural education settings was full of challenges in terms of students' competency, environmental and socio-cultural differences, and practical usability. 

\subsubsection{Competency Differences Between Remote and Local Students}
As presented in Section~\ref{sec:findings-procedure-interaction}, the lecture co-preparing process was a somewhat formalistic form of collaboration because there were so many rural teachers, and it was impossible to balance all requirements. Consequently, teachers in CD schools designed course content based on their own students' abilities. 
Yet, there existed a significant disparity in competency between their students and local students. In local (rural) areas, the historical lack of educational resources and relatively low education quality had resulted in an overall lower competency level of local students. This led to a challenge where course content designed for remote students may not be readily accepted and utilized by local students, which was the major challenge reported by all our participants, including both teachers and students. This challenge was particularly pronounced in science courses, as one student we communicated with mentioned: ``\textit{For humanity courses, it is manageable. However, the teachers speak incredibly fast for natural sciences like Physics, Chemistry, and Mathematics.}''

The differences in students' competency bring about distinctions in teaching content and educational objectives. Our participants expressed that the teaching content of LSRC was more extensive and complex for local students, which was challenging for local students to understand. As P12 noted, \textit{``Their students are highly proficient, and their instruction is comprehensive, encompassing almost all aspects of the subject matter. However, it is hard for our students. Given our students' disadvantage in their foundation training, we sometimes omit more complex topics or cover fewer challenging points in less detail. ''}

Meanwhile, due to the differentiated educational foundation, teaching objectives between remote and local students were also different. \textit{``Teachers at CD School prioritize developing students' thinking abilities, not focusing as much on basic knowledge, assuming their students have already mastered it. But our students need''}, as P9 said. Thus, our participants suggested that the LSRC paradigm was only effective when students were on a similar educational level. 
Additionally, there were significant differences in students' self-discipline and learning motivation: rural students often exhibit lower levels of willingness to learn, initiative, and motivation compared to urban students. Therefore, merely providing resources for autonomous learning was not enough. 



\subsubsection{Environmental and Social-cultural differences Between Remote and Local Settings}
Remote classroom settings were typically in urban or developed resources, where the overall educational atmosphere and parental awareness about education were generally high. However, local settings were often in rural villages with low-income populations, exhibiting significant disparities in educational atmosphere and awareness compared to remote areas. The societal attitude towards education and the level of parental involvement tends to be lower in these regions. These gaps significantly affected local students' adoption of high-quality educational resources, and a lack of parental involvement also presented challenges for such initiatives. As P7 said,

\begin{quote}
    \textit{``…The majority of our area is rural, and there is a significant lack of educational awareness among parents. Many are not cooperative, leaving their children at school and considering education solely the teacher's responsibility. Some parents feel helpless due to their inability to educate, as they are occupied with earning a living and supporting their families…''} (P7)
\end{quote}

Furthermore, differences in societal educational consciousness and atmosphere widened the gap in students' abilities. For instance, rural areas with higher educational consciousness might introduce LSRC earlier, such as in junior high, thus laying a better educational foundation for local students. Consequently, they face fewer challenges in accepting and utilizing LSRC in high school. However, the schools we observed only started this model in high school, making it more difficult for students to adapt. 


\subsubsection{Challenges in Supporting Quality Teaching}
In addition to differences in the educational foundation and competency of rural students and the overall societal awareness of education, the third category of challenges identified by our study stemmed from ICT itself. Specifically, the challenges arose from the live-streaming teaching format, which faced difficulties in adequately supporting personalized education and interaction, as well as guiding students in actively engaging in critical thinking. Specifically, due to the practical LSRC being a one-to-hundreds model, it became impractical to make differentiated adjustments based on the specific characteristics of local students in the teaching process. P3, for instance, compared his actual teaching process with the LSRC teaching process to illustrate this limitation: ``\textit{When we teach in person, we observe the students' reactions. Sometimes, when students appear confused, we would pause and adjust the pace, slowing down to give them time for reflection. However, remote teachers can only learn from their own students.}''


Meanwhile, how to engage students in LSRC class and encourage their independent and active thinking was also a significant challenge reported by our participants. Our participant suggested that it was hard for a "screen" to keep students focused. As P14 expressed, ``\textit{students tend to lose focus while listening. At times, the teacher has already covered several pages, but their textbooks are still a few pages behind}.''

In addition, our participants also considered the LSRC paradigm only addressed the process of "learning" and did not address the process of "practice" or "reinforcement", which were also important for effective learning. As P6 said, ``\textit{In most cases, students can understand the concepts, but after listening, they still cannot solve the problems and are completely unable to apply them – absolutely no bluffing}.'' This challenge was more pronounced in courses that required reasoning, such as mathematics, physics, and chemistry. As P5, a mathematics teacher, said:

\begin{quote}
    \textit{``The main issue is that the derivation process is very complex, especially with those computations that involve a large amount of data and long strings of symbols. They just write it on the board, but students cannot understand it. After listening, students find it almost impossible to tackle problem-solving. In practical application, there are many aspects that need to be addressed, far from the simplicity presented on slides, where a straightforward step leads directly to a result.''} (P5)
\end{quote}

\subsection{Local Teachers’ Efforts for Effectively Integrating Live-Streaming-Based Classroom into Local Classroom}
\label{sec:findings-localteacher}

\begin{figure}[h]
  \centering

  \begin{subfigure}[b]{0.493\textwidth}
    \centering
    \includegraphics[width=\textwidth,height=1.5in]{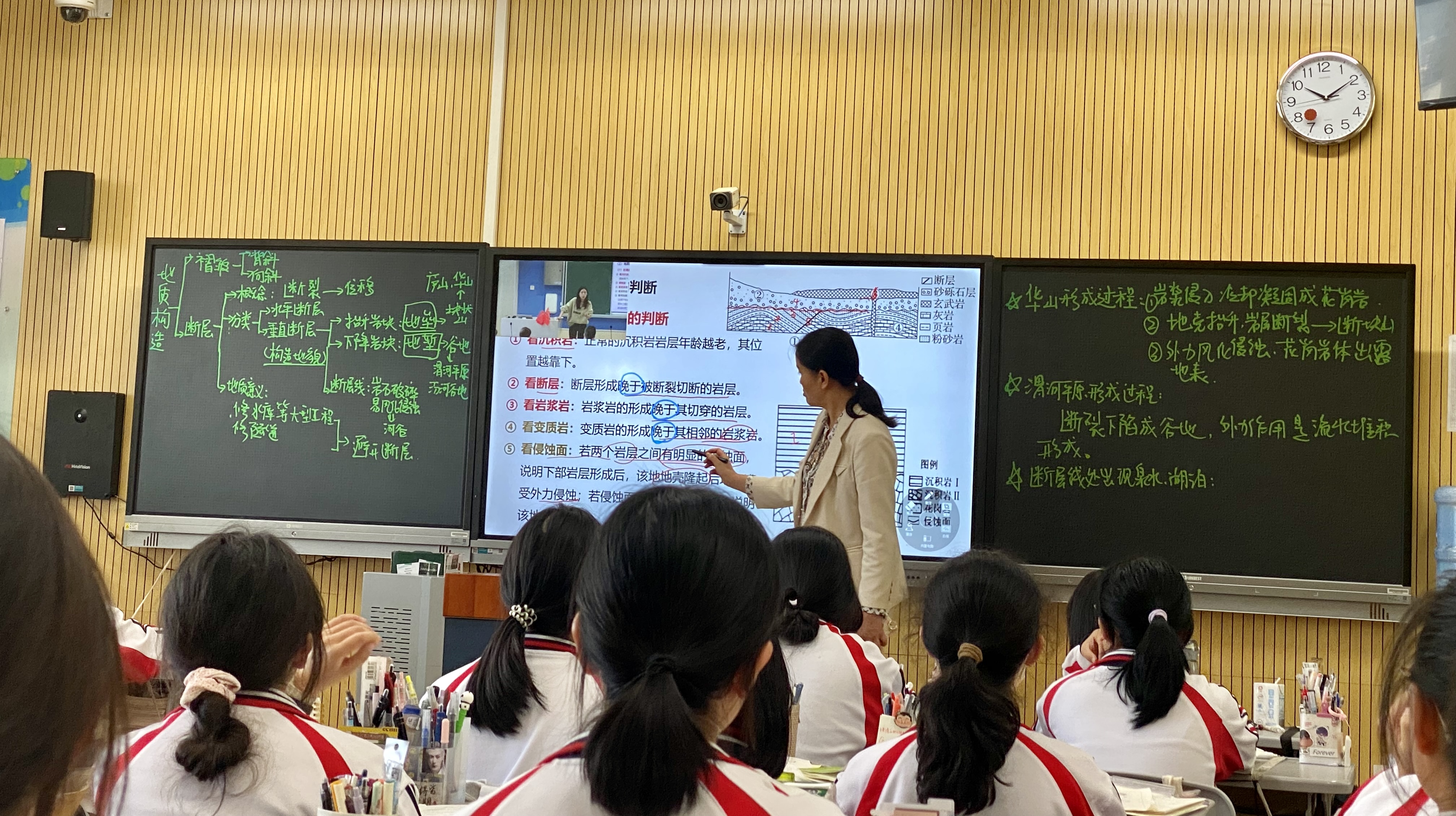}
    \caption{}
    \label{fig:subfig1}
  \end{subfigure}
  \hfill
  \begin{subfigure}[b]{0.493\textwidth}
    \centering
    \includegraphics[width=\textwidth,height=1.5in]{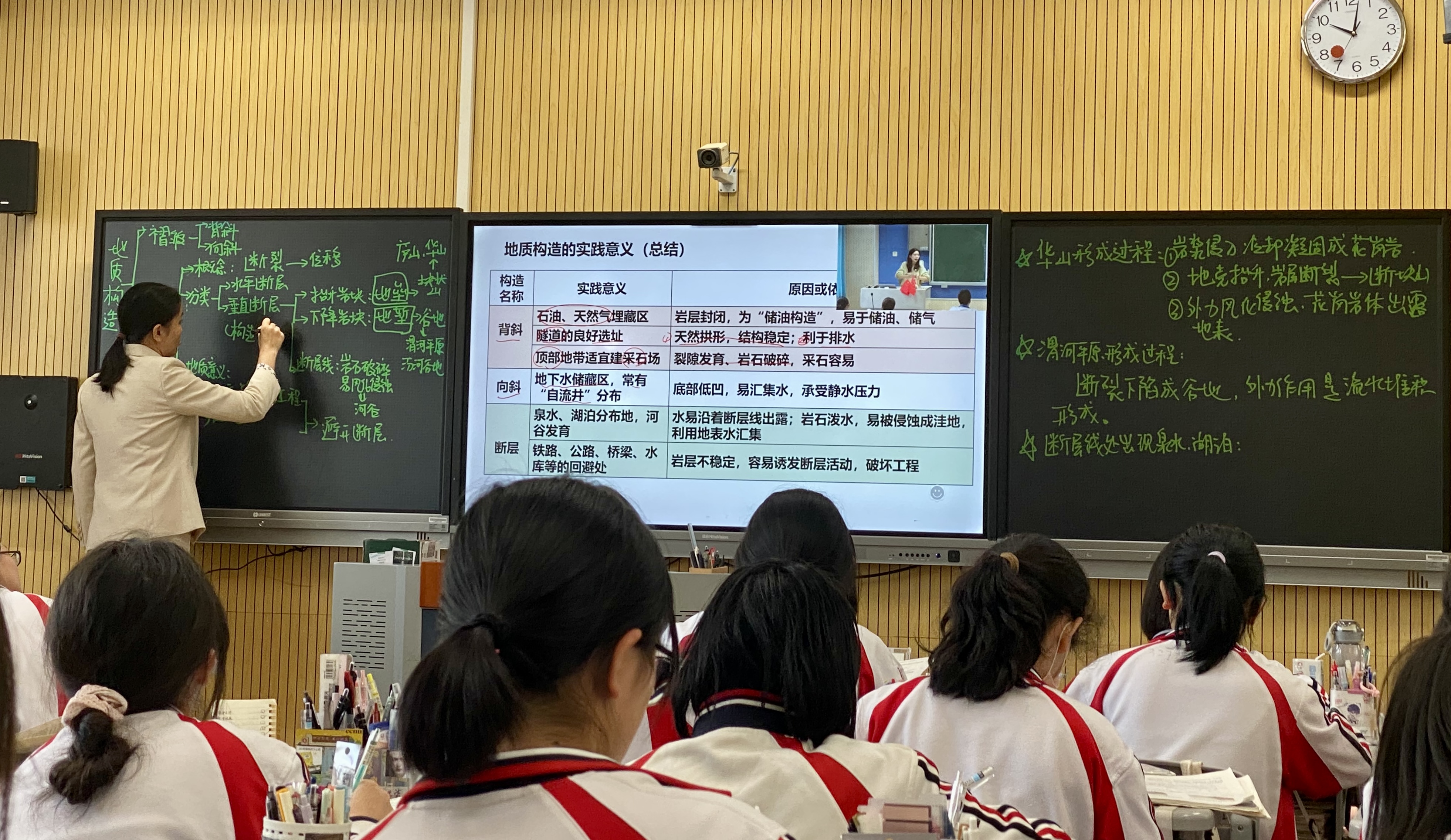}
    \caption{}
    \label{fig:subfig2}
  \end{subfigure}
  
    \medskip
    
  \begin{subfigure}[b]{0.493\textwidth}
    \centering
    \includegraphics[width=\textwidth,height=1.5in]{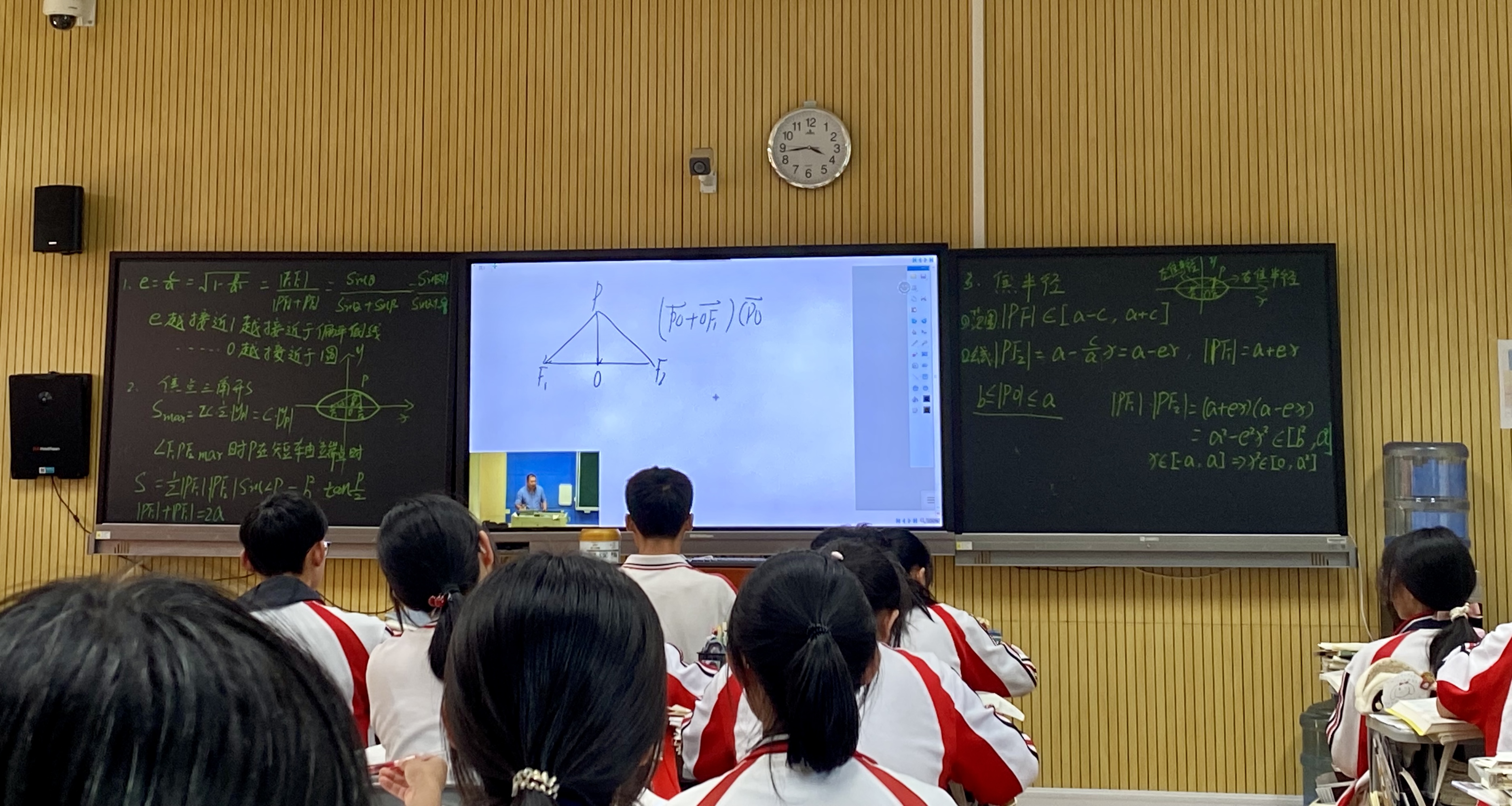}
    \caption{}
    \label{fig:subfig3}
  \end{subfigure}
    \hfill
  \begin{subfigure}[b]{0.493\textwidth}
    \centering
    \includegraphics[width=\textwidth,height=1.5in]{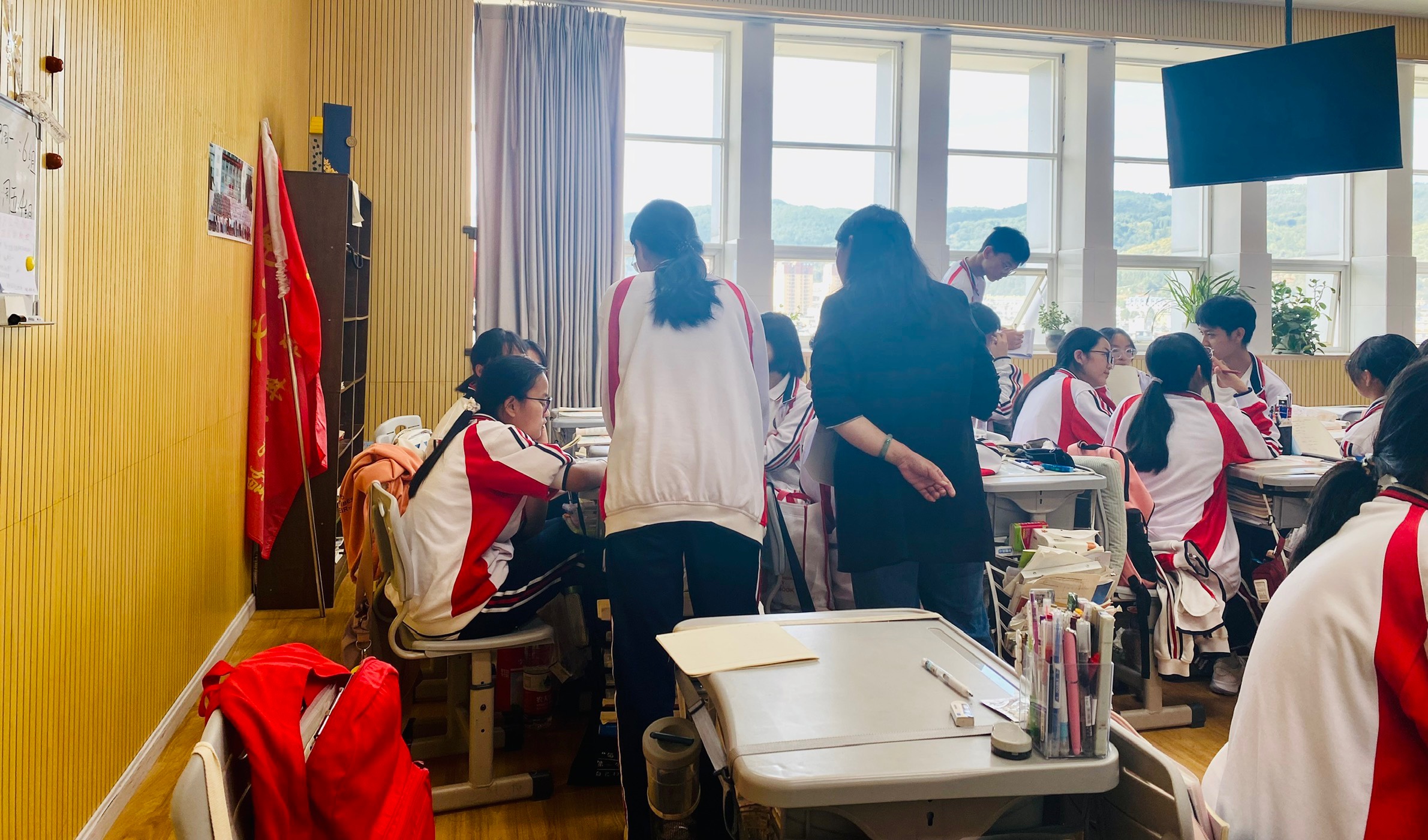}
    \caption{}
    \label{fig:subfig4}
  \end{subfigure}
  
      \medskip
      
  \begin{subfigure}[b]{0.493\textwidth}
    \centering
    \includegraphics[width=\textwidth,height=1.5in]{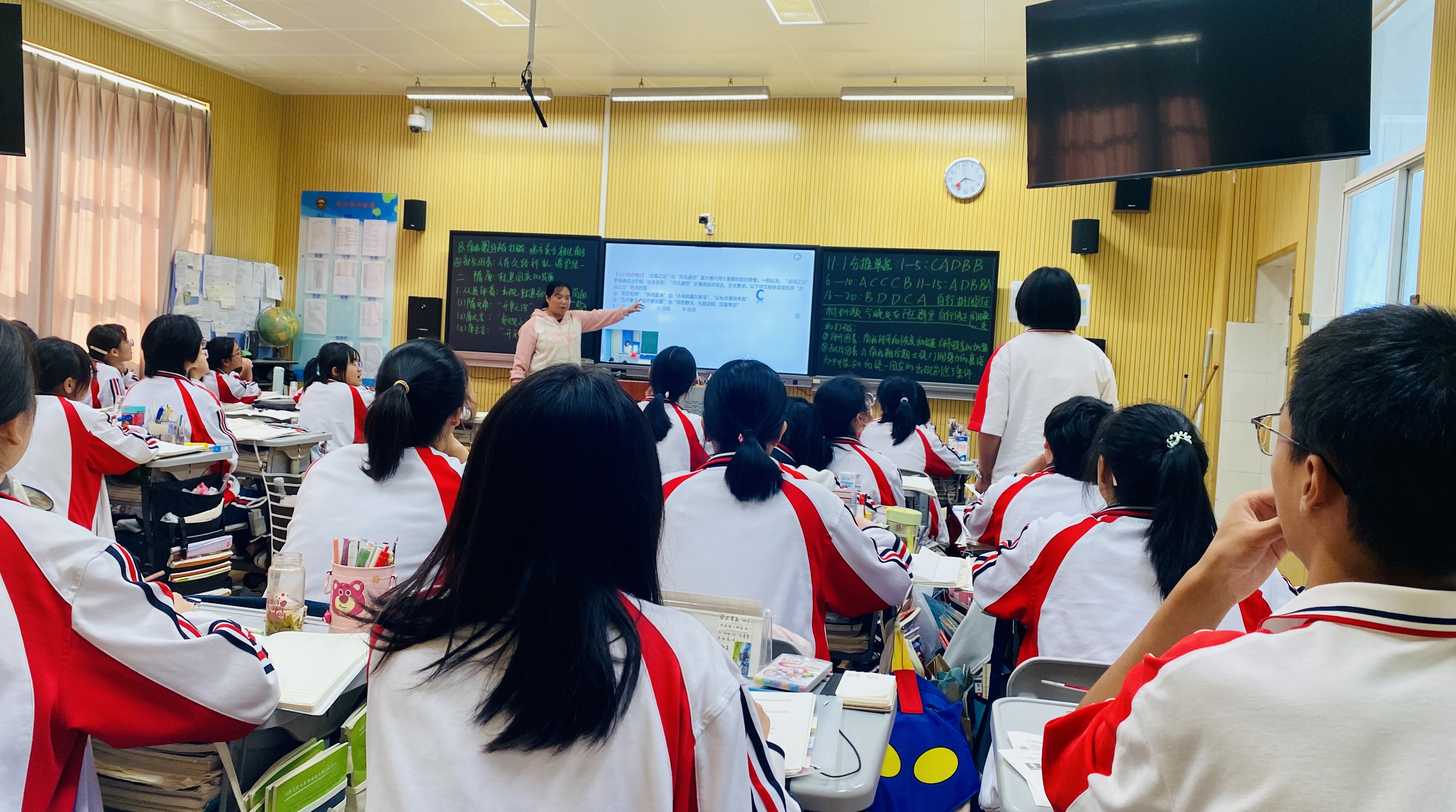}
    \caption{}
    \label{fig:subfig5}
  \end{subfigure}
    \hfill
  \begin{subfigure}[b]{0.493\textwidth}
    \centering
    \includegraphics[width=\textwidth,height=1.5in]{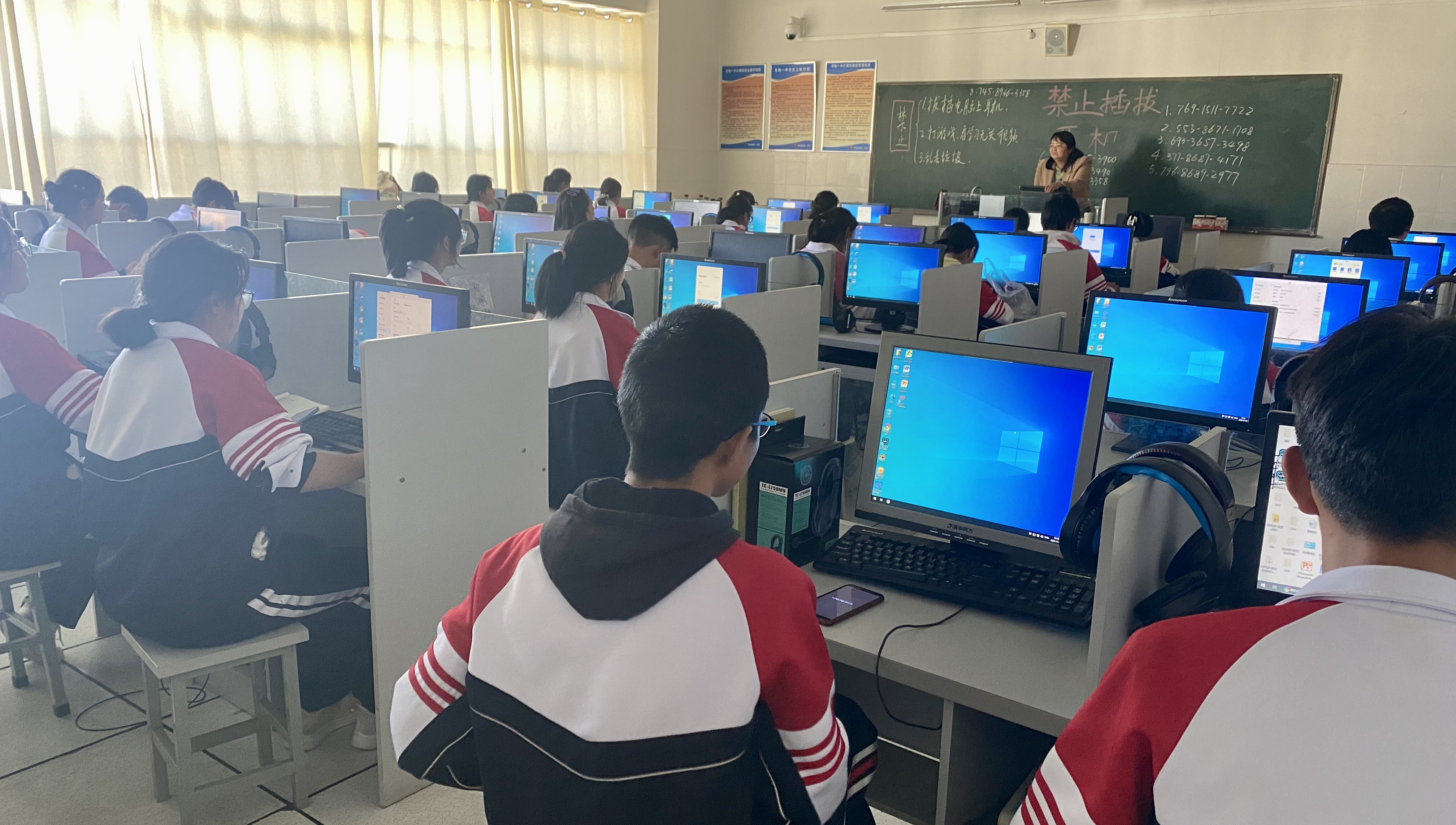}
    \caption{}
    \label{fig:subfig6}
  \end{subfigure}

  \caption{Several LSRC class scenarios. (a) is a geography class, where the local teacher uses a pen to guide students' attention to specific content and highlights key points on the presentation slides. (b)  showcases the geography teacher outlining the class's knowledge structure on the local blackboard. (c) is a mathematics class, with the math teacher writing down essential formulas on the local blackboard. (d) is a physics class wherein the local teacher is organizing group discussions. (e) is a history class, with the teacher interacting with students. (f) is a non-LSRC English class organized by the local teacher to enhance students' English listening skills.}
  \label{fig:total}
\end{figure}

Given the challenges elaborated in Section~\ref{sec:findings-challenges}, our study found that local teachers played an essential and unplaceable role in addressing these challenges and improving the practical effectiveness of LSRC in rural classrooms. 
They, reconfiguring their role from content creators to content delivers, conducted a series of efforts to make the LSRC work effectively and appropriately. From the local students' perspective, all students we informally communicated with also agreed that the role of local teachers was significant for their adoption and understanding of lecture content. We now elaborate on local teachers’ efforts to integrate LSRC effectively into their local classrooms, assisting students' adoption, understanding, and application of lecture content, and improving students' interaction and classroom engagement. 



\subsubsection{Assisting Students’ Adoption and Understanding to Class Content}
Section~\ref{sec:findings-challenges} mentioned that due to differences in students' foundational knowledge and competency, they encountered a series of challenges when adopting and understanding LSRC course content. To address this issue, local teachers reconfigured their teaching practices and conducted many targeted teaching behaviors to help study, adopt, and understand the class content. 

\paragraph{\textbf{Before Class: Preparing the Lessons Meticulously and Guiding Students to Review Relevant Materials}}

Before each class, after obtaining course materials from remote teachers, such as PowerPoint slides and post-class assignment materials, local teachers started their meticulous and time-consuming lesson preparation. P7 described this experience as: 

\begin{quote}
\textit{``You might think our lesson preparation pressure is reduced since we are not lecturing. It is quite the opposite. We prepare even more thoroughly than before. It is essential to anticipate where students might face challenges and prepare necessary materials from textbooks and supplementary sources. For instance, if students struggle with certain words or sentences, laying the groundwork beforehand is crucial, or they will have difficulties during the class.''} (P7)
\end{quote}

Meanwhile, based on the course content, teachers also needed to focus on students' preparation, guiding them to review relevant materials, or else students might not even be aware of the teacher's progress during the class. As P9 commented, \textit{``We must ensure students are well-prepared before class, directing them on topics to review. Without this preparation, they cannot keep up with what's being taught during the class.''} In addition, our participants expressed that this process couldn't be replaced by simply providing students with PowerPoint presentations because \textit{``students might feel they have already seen the material, reducing their engagement in class.''} (P10)

Furthermore, some LSRC classes were designed assuming students have prior knowledge. However, due to differences in middle school curricula, local students might not have learned certain knowledge during their middle schools. In these cases, local teachers needed to fill in gaps from middle school. Figure \ref{fig:total}-f, for instance, showed P7, a first-grade English teacher, was guiding first-year students in practicing their English listening skills, helping them understand the LSRC classes delivered entirely in English.
P7 explained her insights: \textit{``There is a lot of knowledge our students did not learn in middle school, so we need to cover that before class. Sometimes I feel like I am both a high school teacher and a middle school teacher at the same time.''} .

\paragraph{\textbf{During Class: Providing Assisted Instruction and Knowledge Opportunistically}}
In the classroom, due to issues such as students falling behind in thinking, not understanding, or becoming distracted, local teachers needed to conduct certain assisted instructions to address these problems and enhance students' classroom efficiency and participation. As P5, one Math teacher described,

\begin{quote}
    \textit{``During their thinking time, I make the most of it by guiding students through the analysis of the thought process. There is not much time for our students to think on their own. I directly provide some guidance and a pathway for them to follow. Students will not understand anything further if I do not provide cues or guidance. By giving them a small quantity of direction, students have some preparation in advance. When they continue listening, they are more prepared, knowing from which perspective to start thinking.'' } (P5)
\end{quote}

In addition to assisted instruction, most teachers would write supplementary knowledge points, key information, essential words (in English classes), critical formulas and their derivation process (in math classes), and the structure of course knowledge on the blackboard (as shown in Figure 2-a, b, c) to assist students in understanding.
For instance, in English classes, as per the English teacher’s explanation, there was a significant gap in English proficiency between their students and remote students. Most students initially struggled to understand the English lectures by remote teachers. To address this, they had students preview the course content and memorize vocabulary beforehand on the one hand, and wrote tricky words and their Chinese meanings on the blackboard during the lecture to assist in understanding and following along on the other hand. As the English teacher P8 explained, 
\begin{quote}
    \textit{``…When I find some classmates confused, or when a vocabulary may exceed their current vocabulary range, I write them on the blackboard, adding the Chinese translation, to ensure that students can at least understand. This is necessary because they need to understand first. Otherwise, they might get stuck on the vocabulary, completely unable to comprehend.''} (P8)
\end{quote}
 
In addition, some teachers have unique strategies to help students understand. For example, in one geography class, the geography teacher listed course notes on the blackboard (as shown in Figure 2-b)
while lecturing, with the aim of ``\textit{helping students understand the course structure}'' and ``\textit{teaching them how to take class notes}''. She also provided guidance and annotations on the screen (as shown in Figure 2-a), following the remote teacher's content, with the aim of ``\textit{helping students know how far the lecture had progressed and where the key knowledge was}''.

\paragraph{\textbf{After Class: Re-lecturing Key Content, Personalized Tutoring, Dispelling Doubts, etc.}}

After class, teachers provided after-school tutoring on that day’s content, to address students' queries and reiterate key points. As reported by our participants, the content of one class often required half to a full class session for additional explanations. 
\begin{quote}
\textit{``They (remote teachers) teach at a fast pace, cover a wide range of knowledge, and delve deeply into topics. What they cover in one class equals what we cover in four. Their students might grasp it in one session, but ours need four, and still, some struggle. After attending a class, we need another one to two evening sessions to explain the details. We dedicate all our evenings and weekends to this.''} (P2)
\end{quote}

During after-school tutoring, teachers would re-produce the course content, tailoring it to their students' characteristics. As P6 said, \textit{``The original content is designed for \textbf{their} students, and much of it doesn't apply to ours. I have to reassemble and adjust it to suit our students, a task for local teachers. We consider how to present it so our students can better understand and accept it."} Most teachers and students expressed that this process was crucial and couldn't be replaced by re-watching recordings; even recordings were available, and students could review them anytime. The students told us they only watch the recording occasionally, preferring to ask local teachers directly: ``\textit{The video content is good, but our teacher explains it in a way that is more relevant to our practical experience.}''
Teacher P3 explained that from teacher's perspective: \textit{``Asking students to re-watch recordings is not very effective. Our students have foundational gaps, and recording might not help. Even they watch it again, they may not understand. they need further explanation.''} 




\subsubsection{Guiding Interaction and Promoting Student’s Engagement and Proactive Thinking During Class}
During classes, student engagement and active thinking were crucial for improving learning performance. However, as shown in Sections 4.1.2,
and 4.2.4, On one hand, there was limited interaction between remote teachers and local students; on the other hand, the passive listening approach alone was not effective in keeping students' attention and encouraging them to think actively and critically. To address these challenges, local teachers have adopted a series of supplementary interactive methods to enhance students' classroom engagement and guide them in thinking proactively. 

For instance, when a remote teacher asked a remote student a question, the local teacher also prompted their own students to answer questions (as shown in Figure 2-e). Sometimes, local teachers would repeat the questions or instructions from remote teachers, adding interaction with local students to increase their engagement.
When remote teachers organize group discussions or call for answers, local teachers also assist in organizing local group discussions (as shown in Figure 2-d) or have local students stand up to answer questions. In this process, some teachers even muted the voices of the videos, organizing local students to think by themselves. As P15 highlighted: \textit{``They need to think and discuss by themselves; otherwise, they will just be consuming the content fed to them by teachers.''}

Additionally, as presented in Section~\ref{sec:findings-challenges}, local students often exhibited differences in learning enthusiasm and initiative compared to remote students. Their awareness of learning might not be as strong; therefore, local teachers needed to ``\textit{keep an eye on the students}'' in real-time, or else they might become ``\textit{lazy}'' or ``\textit{distracted}''. In our observational study, local teachers paid close attention to students' engagement during class. When they noticed a student appearing distracted, they would directly remind the student or actively encourage her/him to answer questions.

\subsubsection{Promoting Student’s Utility and Application to Knowledge}
We also found that students struggle to truly absorb classroom knowledge by merely listening. Local teachers then bridged this gap to help students learn how to use it, as P7 purposed,
\begin{quote}
\textit{``For most students, a significant portion of the content needs additional explanation and supplementation. Without it, they are still clueless about applying the knowledge or solving related problems. It is like encountering something valuable. It may be excellent, but you still need to familiarize yourself with it and learn to utilize it. All these aspects are vital, and simply listening once is not enough. It is like hearing a novel; you will not remember it after just one listen, let alone such complex material, right?''} (P7)    
\end{quote}

Moreover, local teachers guided laboratory sessions linked to live-streamed courses like physics and chemistry labs. In these sessions, local teachers correlated classroom content with experimental and application processes, enhancing students’ understanding, usage, and application skills of the knowledge.

\subsection{Impacts of Live-Streaming-Based Dual-Teacher Class}
\label{sec:findings-impacts}
Our research suggested that the integration of LSRC into rural classrooms significantly influenced the teaching and learning experiences of local teachers and students. Beyond behavioral changes, it also had substantial impacts on students' psychological well-being, teachers' workloads, self-identity, and more. 

\subsubsection{Impact to Local Teachers: Practices, Workload, Self Identity, and Personal Development}
The impacts of LSRC paradigm on local teachers were primarily reflected in their teaching practices, workload, self-identity, and personal development. In Section~\ref{sec:findings-localteacher}, we have elaborated on how local teachers reconfigured their work practices to integrate LSRC into local classes effectively. Following this, we now discuss how LSRC affects local teachers’ workload, self-identity, and personal development.

\paragraph{\textbf{Reduced Workload? No}}
Under the LSRC paradigm, local teachers' workload appeared to decrease as they were not required to prepare course content or deliver lectures themselves. In many remote live-streaming-based remote class cases, local teachers were often seen as teaching assistants (TAs). However, our study showed that in LSRC-integrated rural contexts, the workload of local teachers had not been reduced; instead, it may even be increased because they needed to perform a series of teaching practices (as reported in Section~\ref{sec:findings-challenges}) to make the LSDC paradigm work effectively. As M15 expressed: 
\begin{quote}
    \textit{``Initially, many people had the idea that if you have a screen, our teachers don't need to do much. They just sit there, and oversee the students listening to remote classes. Even if the expected outcomes were to be realized, many teachers would be unemployed. However, it's not really like that. The reality is that this paradigm has increased the workload for local teachers.''} (M15)
\end{quote}

Our observations confirmed the principal's statement. At XD School, the local teachers' work schedule is extended from 7:00 AM to 11:00 PM every day, seven days a week. The daytime hours from Monday to Friday were for live-streaming classes, while evenings and weekends were for in-depth explanations and tutoring. P11 told us that he had 45 classes per week, leaving little to no time for breaks.

In our interviews with local teachers, they elaborated on the increased workload. Specifically, during the pre-class preparation stage, they needed to be more meticulous, following the curriculum lesson plans, anticipating challenging aspects of the class content, preparing necessary supporting materials, and assisting students in their classroom learning. Particularly, as remote teachers often finalized the course PPT only a day in advance, local teachers typically received the PPT from remote teachers around 5 PM the day before the class and completed these detailed pre-class preparations in the evening.

In class, although teachers didn't need to deliver lectures directly, they had to actively engage, provide guidance where necessary, record the content that required further explanation for later tutoring, and observe students' real-time learning situations. They also interacted with students and enhanced participation when necessary. As P2 said, 

\begin{quote}
    \textit{``In the classroom, we need to be highly focused and listen more attentively. You need to identify and note where the students are having issues, where it is challenging, or predict where they might encounter difficulties. You also need to observe the students' expressions to determine if they are distracted or having trouble understanding. Make a note of the areas where they don't understand, and revisit those during the evening sessions.''} (P2)
\end{quote}

After class, this increased workload became even more apparent as teachers spent much time explaining and reproducing content. This increased workload was more pronounced in science classes (such as Math, Physics, Chemistry, etc.). The science teachers we interviewed all expressed that the remote teachers spoke very quickly, and the content was extensive, making it challenging for many students to keep up. Therefore, more time was needed for after-class explanations. Sometimes, the school adjusts the schedule, allocating evening classes originally intended for liberal arts to science courses, providing science teachers with more time for explanations.

\paragraph{\textbf{Reduced Self- and Social-Identity? No}}
One may speculate that in these teaching scenarios, the role of teachers shifted from content creators and primary educators to teaching assistants, which would impact their self-esteem and social identity. In our study, however, all participants expressed that the format of LSRC had not diminished their sense of self-worth. Instead, they feel proud and confident about their crucial and irreplaceable role in this scenario. Vice Principal M15, who was in charge of this project, expressed: 

\begin{quote}
    \textit{``The role of our local teachers is fundamental. Without their assistance and guidance, relying solely on 'screens' for learning is inadequate. If this LSRC paradigm succeeds, the credit should largely go to them. They contribute significantly to its success. Without their involvement and efforts, this initiative would not be feasible.''} (M15)
\end{quote}

\paragraph{\textbf{Impact to Teachers’ Self Development}}
After the integration of LSRC model, local teachers could directly participate in the teaching process with remote teachers, providing them with opportunities for learning and growth. As P2 mentioned, \textit{``These types of classes are very beneficial for us too. We have learned a lot from the remote teachers. For example, in today's class, the topic of the conic section was complex, but Mr. Li (remote teacher) provided an in-depth explanation. His approach is different from ours; he clarifies the logic for students. If we do not pay close attention, we often would not come up with these ideas independently.''}

Yet, even though local teachers had learned all the courses, all our participants acknowledged that reaching the same level as remote teachers remained challenging due to the local social environment, their own learning mindset, and their busy work schedules. Vice Principal M15 further explained: ``\textit{I think the main issue is teachers' awareness. Everyone's thinking should evolve with the times, and that is something you cannot just learn. Teachers in our small county lack access to new information and resources, and there is no systematic opportunity for professional development. Many teachers are also reluctant to learn proactively. Time is another factor; we are occupied with numerous tasks daily and do not have time to learn. We spend all our holidays teaching classes to the students.}''

\subsubsection{Impacts to Local Students: Beyond Academic Performance}
Existing literature has demonstrated the effectiveness of this approach in improving students' academic performance \cite{bai2023impact, mo2015computer, zheng2015understanding}. Our study echos this conclusion. 
Moreover, our study suggested that the impact of LSRC on students went beyond academic performance to broadening students' horizons and giving them hope and goals. As teacher P3 considered: \textit{``The impact is not just in grades but in their insight and thinking. When their students (remote students) speak confidently on the screen, our students admire, imitate, and learn from them, viewing them as role models. They, therefore, have something to aspire to.''} 


Regardless of the positive impact, our study found that LSRC also imposed significant psychological pressure on many students as they saw the vast gap between themselves and urban students. They need to spend much more time trying to catch up, and sometimes it seems unattainable. One teacher (P4)comments, \textit{``If there were no exams, it would be very positive. However, exams are quite discouraging. There is certainly a psychological gap, especially for these students, who have been at the top in their school from elementary to middle school, always leading. Coming here and finding themselves trailing after an exam can be quite disheartening. Some experience a significant setback.''  }  

While acknowledging the stressful situation, particularly in the beginning, the majority of the students had positively embraced this gap and consistently worked hard to bridge it. As one student said, ``\textit{This is a new teaching method, which, for us, is both an opportunity and a challenge. It demands a high level of individual capability, requiring self-adjustment, which is crucial.}''






  
\section{Discussion}
\label{sec:discussion}
In this paper, we have presented a qualitative study of live-streaming-based remote classrooms (LSRC) in rural China. Our findings suggest that while LSRC enables rural students equal access to high-quality educational resources, its practical implementation faces numerous challenges. In particular, we emphasize the crucial role of local teachers in addressing these challenges, ultimately achieving the desired improvement of students’ learning outcomes. We also examine the impact of LSRC on the original education ecosystem in rural areas. We now reflect on our findings and discuss the implications for future ICT-mediated technology interventions in rural educational settings.

\subsection{Recognizing the Significant Role of Local Teachers}
\label{sec:discussion-roleoflocalteacher}
Under the promised live-streaming-based education paradigm for rural areas, local teachers were often identified in a disadvantaged or overlooked role \cite{mitra2005acquisition, ames2016learning, isci2015use}. In these similar research cases, the identity of local teachers has been directly replaced with that of teaching assistants (TAs), primarily serving as instructional aides. The increasing potential of educational AI has further ignited discussions about the future role of teachers, with some viewpoints suggesting that the demand for live and human teachers will diminish \cite{selwyn2019should}, or that they may become obsolete, as AI applications continue to improve \cite{rovzman2023building}. This perception is also reflected in our study. That is, the role of local teachers has not received sufficient attention. Most mainstream media in China coverage of this phenomenon (e.g.~\cite{destiny, destiny2}) highlights the positive impact of this paradigm on students' destinies, attributing this positive influence to the development of ICT, high-quality teaching resources, and advanced teaching methods from remote teachers. However, the efforts of local teachers have been notably absent from the current mainstream media narratives. Our research challenges this perception. We argue that local teachers play a crucial role in enabling the effectiveness of various technological interventions in rural educational settings. This significance is primarily manifested in several key aspects.

Firstly, due to sub-optimal educational foundation, rural students face significant challenges in adopting, understanding, and applying the delivered high-quality educational resources. To address these, local teachers reconfigure their teaching duties, focusing on assisting students in understanding and applying knowledge. They are not the commonly perceived teaching assistants (TAs) solely assisting remote teachers, who may not need much course preparation. On the contrary, they engage in almost all interactions with local students, as indicated in Figure 1. They still need course preparation, and in fact, their practical preparation process is more meticulous. They need to understand the teaching logic of remote teachers, anticipate challenges local students may encounter in assimilating the content, and prepare necessary supplementary materials to address these challenges. After class, they further explain the course content to ensure better understanding by students. All participants, whether teachers, students, or administrators, universally agree that the positive impact expected cannot be achieved without the involvement of local teachers.

Secondly, the promise for the effectiveness of high-quality educational resources or intelligent educational tools in supporting students' independent learning is that students have sufficient motivation and self-discipline to engage in proactive learning \cite{chiu2023teacher}. However, many students lack such strong self-motivation and self-restraint abilities \cite{espinosa2015self}, particularly in rural areas \cite{de2014reasons}. They need a person (i.e., a teacher) to provide guidance.
Moreover, the critical role of local teachers is also reflected in their attention to students' psychological well-being and their daily routine management, i.e., the aspects that AI cannot replace.

In existing CSCW and HCI literature, discussions on the impact of technology intervention on the role of teachers have been only reported recently. For instance, \citet{guo2022remote} et al. reported the roles of local teachers in remote co-teaching programs in making indirect impacts on their students by deeply coordinating with remote teachers, as well as the challenges encountered by local teachers such as lack of instruction, training, rewards, and stimulation. \citet{varanasi2019teachers} et al. reported how teachers in India reconfigure their work practices, including course preparation, in-classroom teaching practices, bureaucratic work processes, and post-teaching feedback mechanisms, around a teacher-oriented technology intervention. We commend such research efforts and suggest that in future design and implementation of technological interventions for rural educational settings, the focus should not only be on the designs targeting students' learning performance and how these interventions improve students' academic outcomes and change their destinies. It should also consider the impact of these technologies on teacher instructional practices, behaviors, identity, and self-worth. In the value evaluation system, the crucial efforts of rural teachers in this process need to be recognized and acknowledged. Through this scrutiny, we can design more humanistic educational technologies in which human (local teachers) value and significance could be better recognized, respected, and supported. We believe such efforts will help improve the workplace well-being of local teachers, and alleviating the global issue of a shortage of rural teacher resources.

\subsection{Enabling Social Ecology}
\label{sec:discussion-socialecology}
We have demonstrated that various social and ecological factors significantly impact the usability and effectiveness of live-streaming-based technology interventions in practical rural educational settings. This influence is multi-dimensional. First, the local socio-cultural and educational consciousness directly determines students' foundational knowledge and learning abilities, which in turn affects their acceptance and utilization of high-quality education resources. Second, although teachers have been shown to play a crucial role in the actual effectiveness of this education paradigm to students, the positive impact of this model on the personal development of local teachers is not directly reflected. In the context we studied, local teachers spent almost all their time on the demanding task of delivering knowledge, leaving little time for improving their own professional skills. Particularly, our participants did not even exhibit a strong awareness of self-improvement. This is closely related to the educational ecology and social consciousness in local areas. Many inherent social stereotypes about rural backwardness are deeply ingrained in their minds. They feel that they cannot reach the same level as remote (urban) teachers, even if they are proficient in all course content and better understand local students. This restricts the broader application and promotion of this paradigm, as well as the optimization of the local educational ecology.

Research in CSCW and related fields has often focused on analyzing and supporting practices, interaction, and collaboration in ICT-mediated educational models, e.g., the interaction and collaboration between remote and local teachers in an ICT-mediated co-teaching paradigm \cite{guo2022remote}, the adoption and usage of teachers and students to live-streaming-based \cite{chen2021Was, chen2021learning} and MOOC-based online education \cite{zheng2015understanding}, etc. However, building on prior work in HCI on `socio-technical systems \cite{baxter2011socio}' and `situated actions \cite{suchman1987plans}', we argue that the examination focused solely on interaction and collaboration cannot fully explain how individuals creatively and collectively assemble resources to achieve their goals in a specific social and cultural situatedness, or how they encounter the value systems and sociopolitical realities that are part of their engagement with technology. 

Based on our findings, we emphasize the importance of enabling social ecologies when designing and deploying technology interventions in rural educational settings. While previous research has acknowledged the specific social and cultural characteristics of rural areas (e.g.\cite{de2014reasons, johnson2007rural, guzman2021dropout}), most studies treat these features as problems or attributions to poor rural education system, aiming to solve them through technological solutions. Rather, our research considers these social and ecological features as the social reality of rural areas, which educational technology intervention design should draw on.

By deploying technology interventions within enabled social ecologies, we can anticipate a more sustainable and scalable paradigm for educational technology interventions and the improvement of education quality at the societal level.
For example, we can elevate the "one screen for all" live-streaming model to a multi-tiered dissemination model. At each tier, the local educational department would localize and adapt the upper-tier education content and resources to better align with the social and ecological context of local students. This approach might enhance the relevance of course content to local students and improve their adoption and understanding on the one hand, fully utilize and disseminate the delivery process and knowledge of local teachers, significantly reducing local teachers' workload and improving teaching efficiency. This social ecology-enabled multi-tiered dissemination model can empower more local students, thereby contributing to the educational advancement of the entire region.

\subsection{Education Quality, Equality, and Equity}
\label{sec:discussion-qualityequalityequity}
Undoubtedly, technological interventions have tremendous promise and potential for addressing the cross-region educational inequality and improving the quality of education in economically underdeveloped, rural and isolated areas, through, for instance, providing higher-quality educational resources to students there \cite{mitra2005acquisition, ames2016learning, techred, yikuaiping}. However, echoing the calls made by existing literature \cite{sulecio2018avoiding, yang2018promoting}, our study also advocates for a more thorough examination of this promise, ensuring that the implementation of technological interventions aligns with our expectations. We now turn to a more macro perspective to discuss the impact of technology interventions on education, through the lens of educational quality, equality, and equity.

\textbf{Education Quality: More Exam-Oriented Education?} In our research case, we found that under the live-streaming-based educational paradigm, both teachers and students dedicate almost all of their time, including evenings and weekends, to learn, digest, and absorb this high-quality course content and knowledge. They engage in learning from 7 a.m. to 11 p.m. daily, seven days a week, with only two days of rest per month. The daytime of weekdays (Monday to Friday) is devoted to participating in live-streaming courses, while evenings and weekends are utilized for supplementary classes and in-depth understanding of the content. Both students and teachers operate at an exceptionally fast pace of learning and working.

This raises a concern: could this trend lead to an education system increasingly focused on exam-oriented learning? While short-term improvements in academic performance are unquestionable, other essential aspects of holistic education at the foundational level are being compressed, even neglected. We consider this to be a potential negative impact that this kind of technology intervention may bring about when integrating into rural and isolated settings. A possible solution could be to implement this intervention earlier, starting perhaps from middle school, enabling students to establish a solid foundation at an earlier stage, as opposed to the case we studied, which commenced at the high school level. This requires concerted efforts from policy decision-makers.

\textbf{Education Equality and Equity: Improved or Aggravated?} One Screen Initiative in China \cite{yikuaiping}, as well as many other initiatives (e.g., \cite{mitra2005acquisition, ames2016learning}), were first conceived with the goal of bridging educational gaps and enhancing educational equality and equity. Through an empirical study of One Screen Initiative, we consider that the intervention of live-streaming-based classes has indeed, to some extent, addressed the issue of educational disparity, enabling more students in rural and isolated villages to access high-quality educational resources that were previously only available in advantaged urban settings. However, our findings suggest that this approach has not fundamentally resolved the issue of educational equity, i.e., distributing resources and opportunities based on the specific needs of individuals or groups \cite{twoTerms2}, considering their inherent disadvantage. Even it may exacerbate educational inequity, as not all students can avail themselves of such highly-condensed content with a fast-paced delivery approach.  

In our investigation of XD School, the local premier high school, only a small percentage of classes have adopted this education model, while more students have not, with reasons of "it won't make a difference even if we provide them (live-streaming-based class)." According to our participants, some other schools in the area previously also employed similar classes, but the majority abandoned them in time due to ineffective outcomes. The immediate consequence of this situation is that a limited number of high-achieving students truly benefit from this model, becoming the celebrated "children whose destinies were changed" as seen in news reports \cite{yikuaiping, destiny2}. Simultaneously, a larger number of average students continue to navigate their original paths of learning and living, with the gap widening further between them and their peers who have embraced the new approach.

Our study argues that merely providing access to high-quality resources is insufficient for students in rural and isolated areas. Particularly, as future ICT technologies become more pervasive and affordable, along with the increasing availability of free online educational resources (e.g., \cite{Chinak12eduplatform, USk12eduplatform}), equal access to high-quality educational materials will no longer be the primary challenge facing rural education. The more challenging issue will be how to enable rural students to effectively use these educational resources, achieving the expected goals of educational equity and literacy enhancement. Yet, much of the efforts are still spent on how to deliver higher-quality educational resources to students \cite{makalima2023co, arawjo2019computing, khan2019mitigating}. Our study complements these efforts by illustrating how the design and deployment of technological interventions need to fully consider the actual learning and acceptance capabilities of rural students, as well as their social and cultural environment, instead of "one screen for all." We should focus on technological uptake and effectiveness in practice and realize the expected more equitable educational ecosystem.

\subsection{Design Implications}
\label{sec:discussion-designimplications}
Based on our findings and drawn on the discussion above, we propose a series of practicable technological design recommendations tailored for live-streaming-based education technology to enhance its practical efficacy and effectiveness, especially in rural educational scenarios. 
First, our findings suggest that for rural students, it's not necessarily the higher quality of learning resources that are more beneficial, but rather those that are tailored to their foundational knowledge and capabilities. We therefore suggest differentiating course content based on dimensions such as difficulty, breadth, and depth, allowing local students and teachers to choose learning materials that suit their individual abilities. 
Second, more human-centric assistive functions, such as automatically generating knowledge frameworks based on course content, highlighting key points, and providing necessary prompts for difficult contents, concepts, or words, could be incorporated into existing platforms to aid students in understanding knowledge, enhancing their class engagement and improving learning outcomes. These functions are proposed according to the behaviors our participants are commonly adopting in their live-streaming-based classes, deemed effective in enhancing students’ engagement and learning performance.
Moreover, with the emergence of more intelligent large language models (LLMs), features like LLM-assisted tutoring and explanations may aid students in further understanding sections they could not immediately comprehend during the class.

It's crucial to emphasize that our intention is not to enhance current technological interventions to be more intelligent or to replace teachers. Instead, echoing the existing discussion on the human-computer collaborative relationship (e.g., \cite{ren2019human, ma2018towards}), we aim for the inclusion of these features to alleviate the workload of local teachers, providing them with more time to assist students and focus on self-improvement.

\subsection{Limitation and Future Work}
We acknowledge that, because XD school where we conducted our research predominantly involved relatively young teachers into the live-streaming-based remote classroom project, we didn’t have the opportunities to explore the impact of diverse demographic characteristics and work experiences of teachers on the adoption, usage, and experience of ICT intervention technologies in rural school settings. Meanwhile, our study took place in a specific rural Middle School (XD school), and the schools may not be representative of all Chinese rural schools.
In future research, schools with longer employment of such technology interventions and more diverse socio-cultural and regional characteristics, as well as teachers with more diverse demographic characteristics and work experiences should be investigated to achieve a more comprehensive understanding of the effectiveness and impact of such technology interventions in rural educational settings.
In addition, our findings are based on a specific case study of live-streaming-based classrooms in China. Thus, there is a possibility that some of our findings are specific to this live-streaming-based technology intervention in rural schools. Future work will be necessary to understand how our findings may generalize to other technology interventions in rural settings.

\section{Conclusion}

In this paper, we provide an empirical examination of live-streaming-based remote classrooms (LSRC) in rural China. 
Our findings report a series of practical challenges when integrating LSRC into rural classrooms and how local teachers reconfigure their identity from content creators to content deliverers and enable students to better absorb and comprehend this content through a series of efforts, ultimately achieving the desired improvement of students' learning outcomes.
Building upon our findings, we call for a reconsideration of interaction paradigms and evaluation systems of ICT-mediated rural education, emphasizing the significance of rural teachers. We end by discussing the implications for future ICT-mediated technology intervention in rural settings.

\begin{acks}
To be added.
\end{acks}

\bibliographystyle{ACM-Reference-Format}
\bibliography{sample-base}










\end{document}